\documentclass[10pt]{revtex4}
\pdfoutput=1
\usepackage{amssymb}
\usepackage{latexsym}
\usepackage{epsfig}
\begin{document}

\title{  Emergence of the world with  Lie-N-algebra and M-dimensions from nothing  }

\author{Alireza Sepehri $^{1,2}$\footnote{alireza.sepehri@uk.ac.ir}, Richard Pincak $^{3,4}$\footnote{pincak@saske.sk}}
 \affiliation{ $^{1}$Faculty of
Physics, Shahid Bahonar University, P.O. Box 76175, Kerman,
Iran.\\$^{2}$ Research Institute for Astronomy and Astrophysics of
Maragha (RIAAM), P.O. Box 55134-441, Maragha, Iran.\\$^{3}$ Institute of Experimental Physics, Slovak Academy of Sciences,
			Watsonova 47,043 53 Kosice, Slovak Republic.\\ $^{4}$Bogoliubov Laboratory of Theoretical Physics, Joint
						Institute for Nuclear Research, 141980 Dubna, Moscow region, Russia }

\begin{abstract}
 In this paper, we propose a new model in Lie-N-algebra that removes the big bang singularity and produces the world with all it's objects and dimensions from nothing. We name this theory as G-theory. In this model, first, two types of energies with opposite signs are produced from nothing such as the sum over them be zero. They create two types of branes with opposite quantum numbers which interact with each other by exchanging bosonic tensor fields like the graviton and compact. By compacting branes, fermionic tensor fields are emerged  which some of them play the role of the gravitino. Also, some dimensions take extra "i" factors, their properties become different and they behave like time dimensions. Gravitons and gravitinoes create two types of wormholes which lead to the oscillation of branes between expansion and contracting branches. These wormholes produce a repulsive gravity in compacted branes and cause that their particles get away from each other and expansion branch begin. Also, they create an attractive gravity in opening branes and lead to closing their particles and starting the contraction epoch. Our universe is born on one of these branes and oscillates between contraction and expansion branch.


 \end{abstract}
 \date{\today}

\maketitle
\section{Introduction}
One of the main puzzles in cosmology is the Big Bang singularity and the events that occur before it and lead to the concentration of infinite amounts of energy at one point. Until now, some models have been proposed to get rid of
       the big-bang singularity in string theory and M-theory. These theories are more general respect to usual theories and can solve many puzzles in cosmology. There are various versions of these theories, for example, type IIA, type IIB, type I, heterotic (SO(32)) and $E_{8}\times E_{8}$ that each of them may describe some parts of phenomenological events in cosmology, QCD and other fields of physics \cite{m1}. These theories argued about ten dimensional space-time, however the maximum dimension allowed
       by supersymmetry of the elementary particles is eleven. In 1987, it has been asserted that  the
       Type IIA string may be  the limiting case of the eleven-dimensional supermem-
       brane \cite{m2} and in 1994, it was discussed that the spectrum of states that
       are created by compactifying the membrane theory from eleven dimensions to four
       are the same of  those that derived by compactifying the Type IIA string from ten
       dimensions to four \cite{m3,m4}. In 1995, Edward Witten has suggested a new theory, named M-theory whose low energy effective field theory description is 11-dimensional supergravity \cite{t1}. This theory may be transformed to various string theories by compactification and then by various string dualities. However, the algebra that should be used for M2-branes was puzzle. Less than ten years ago, some authors proposed a formalism for M2-branes and argued that for defining the action with $N=8$ supersymmetry (which is an accepted supersymmetry in eleven dimensions), the Lie-3-algebra is helpful  \cite{h13,h14,h15,h16}. They introduced two form gauge fields, spinors and scalars in M-theory by applying this algebra and calculated the relations between them. 
       
        In models which remove the big bang in M-theory, reverse to usual models in cosmology \cite{h1}, first, there aren't any gauge field and fermion and the world only consists of the scalar fields that are linked to zero dimensional objects, named  M0-branes \cite{h2,h3,h4,h5}. Then, the $M0$-branes join to each other and build
        a system of $M1$ and anti-$M1$-branes connected by a wormhole which is known as an $M1$-BIon \cite{h4}.  When the M0-branes link to each other 
       symmetrically, gauge 
       fields are created and by joining $M0$-branes anti-symmetrically, such as the upper and lower of $M1$-branes are different, fermions are born \cite{h3}. By linking M1-BIons to each other, $M3$-BIons are built, which each of them includes an $M3$, an anti-$M3$-brane and a wormhole that is a bridge between them
       \cite{h6,h7,h8}. Our universe is constructed on one of these M3-branes and by opening them, expands   \cite{h3}. However, M-theory can't create the phenomenological events in four dimensional universe. Because,   only two stable objects, $M2$ and $M5$ are existed in this theory and the life-time of  $M3$-brane which our universe is placed on it, is ignorable. On the other hand, properties of stable branes ( $M2$ and  $M5$) are different from our  universe and we can't apply  these objects for constructing the present universe. In additional, the origin of bosonic and fermionic fields and totally the supersymmetry is unclear in this theory.
       
       To solve these puzzles, we have to build one bigger theory which removes the big bang singularity and considers the process of creation of all things from nothings. To construct this theory, we extend dimensions of algebra from two and three in string theory and M-theory to N and number of time-spacial dimension from 10 and 11 in these theories to M. We assume that  at the biginning, there doesn't exist any energy, field and branes. Then, two positive and negative energies are created such as the sum over was zero. These energies produce M dimensions and two branes with opposite quantum numbers and some bosonic tensor fields. These fields which their rank changes from zero to dimension of branes, interact with each other and lead to compacting of branes. During this compacting, some  extra $i$'s have been added to some dimensions and give the properties of time to them. Also, by compacting branes, fermionic superpartners of bosonic fields are born and supersymmetry emerges. Bosonic fields produce bosonic wormholes which by creation of attractive force, prevent from expansion of branes. Fermions create fermionic wormholes which by production repulsive force, prevent from contraction of branes. Thus, branes oscillate between contraction and expansion branches. 
       
In section \ref{o1}, we will consider  the process of formation of branes  and the emergence of supersymmetry and supergravity in G-theory.  In section \ref{o2}, we will construct BIon and obtain the scale factor of universe and parameters of wormhole in terms of time. In section \ref{o3}, we will consider the origin of Horava-Witten mechanism in G-theory.  The last section is devoted to a summary and conclusion.

The units used throughout the paper are: $\hbar=c=8\pi G=1$.

\section{Emergence of the supersymmetry and the supergravity in G-theory  }\label{o1}
Previously, it has been shown that all Dp-branes in string theories are built from D0-branes which obey from Lie-two-algebra with two dimensional brackets \cite{h2,h9,h10,h11,h12}. Also, all Mp-branes in M-theory are constructed from M0-branes  which obey from Lie-three-algebra with three dimensional brackets \cite{h13,h14,h15,h16}. Now, by using lie-N-algebra and  generalizing  dimensions to M, we construct a new theory which includes all properties of string theory and M-theory and resolve puzzles in them. To show this, we begin of the action for Dp-brane \cite{h9,h10,h11,h12}: 

\begin{eqnarray}
 &&S=-\frac{T_{Dp}}{2}\int d^{p+1}x \sum_{n=1}^{p}\beta_{n} \chi^{\mu_{0}}_{[\mu_{0}}\chi^{\mu_{1}}_{\mu_{1}}...\chi^{\mu_{n}}_{\mu_{n}]}\label{s4}
\end{eqnarray}

\begin{eqnarray}
\chi^{\mu}_{\nu}\equiv  \delta ^{\mu}_{\nu} STr \Bigg(-det(P_{ab}[E_{mn}
E_{mi}(Q^{-1}+\delta)^{ij}E_{jn}]+ \lambda F_{ab})det(Q^{i}_{j})\Bigg)^{1/2}~~
\label{r1}
\end{eqnarray}

where

\begin{eqnarray}
   E_{mn} = G_{mn} + B_{mn}, \qquad  Q^{i}_{j} = \delta^{i}_{j} + i\lambda[X^{j},X^{k}]E_{kj} \label{r2}
\end{eqnarray}

 $\lambda=2\pi l_{s}^{2}$, $G_{ab}=\eta_{ab}+\partial_{a}X^{i}\partial_{b}X^{i}$ and $X^{i}$ are  scalar strings that link  to branes. In this equation, $a,b=0,1,...,p$
are the world-volume indices of the Dp-branes, $i,j,k = p+1,...,9$
are the indices of the transverse space, and m,n are related to
ten-dimensional spacetime indices. Also,
$T_{Dp}=\frac{1}{g_{s}(2\pi)^{p}l_{s}^{p+1}}$ refers the tension of
Dp-brane, $l_{s}$ is the string length and $g_{s}$ refers to the string
coupling. Now, we can show that the acion of Dp-brane can be constructed by summing over actions of D0-branes. To this end, we use of  
the below rules \cite{h2,h4,h9,h10,h11,h12}:

\begin{eqnarray}
&& \Sigma_{a=0}^{p}\Sigma_{m=0}^{9}\rightarrow \frac{1}{(2\pi l_{s})^{p}}\int d^{p+1}\sigma \Sigma_{m=p+1}^{9}\Sigma_{a=0}^{p} \qquad \lambda = 2\pi l_{s}^{2}\nonumber \\
&& i,j=p+1,..,9\qquad a,b=0,1,...p\qquad m,n=0,1,..,9 \nonumber \\
&& i,j\rightarrow a,b \Rightarrow [X^{a},X^{i}]=i \lambda
\partial_{a}X^{i}\qquad  [X^{a},X^{b}]=\frac{ i \lambda F^{ab}}{2} \nonumber \\
&& \frac{1}{Q}\rightarrow \sum_{n=1}^{p}
\frac{1}{Q}(\partial_{a}X^{i}\partial_{b}X^{i}+\frac{\lambda^{2}}{4} (F^{ab})^{2})^{n}\nonumber \\
&& det(Q^{i}_{j})\rightarrow det(Q^{i}_{j})\prod_{n=1}^{p}
det(\partial_{a_{n}}X^{i}\partial_{b_{n}}X^{i}+\frac{\lambda^{2}}{4} (F^{a_{n}b_{n}})^{2}) \label{r3}
\end{eqnarray}

Now, we can show that the action of Dp-branes can be written in terms of two dimensional brackets \cite{h2,h4,h9,h10,h11,h12}:

\begin{eqnarray}
&& S_{Dp} = -(T_{D0})^{p} \int dt \sum_{n=1}^{p}\beta_{n}\Big(
\delta^{a_{1},a_{2}...a_{n}}_{b_{1}b_{2}....b_{n}}L^{b_{1}}_{a_{1}}...L^{b_{n}}_{a_{n}}\Big)^{1/2}\nonumber\\&&
(L)^{b}_{a}=Tr\Big( \Sigma_{a,b=0}^{p}\Sigma_{j=p+1}^{9}(
[X^{a},X^{j}][X_{b},X_{j}]+[X^{a},X^{b}][X_{b},X_{a}]+[X^{i},X^{j}][X_{i},X_{j}])\Big) \label{s7}
\end{eqnarray}

where we have used of the antisymmetric properties for $\delta$. At this stage, we can show that above action is built by summing over actions of D0-branes by using the below definition for  D0-brane \cite{h9,h10,h11,h12,r12,rr12}:

\begin{eqnarray}
&& S_{D0} = -T_{D0} \int dt Tr( \Sigma_{m=0}^{9}
[X^{m},X^{n}]^{2}) \label{s8}
\end{eqnarray}

 which by applying $[X^{0},X^{n}]=\partial_{t}X^{n}$\cite{h2,h3,h4,h5} transforms to :
 
\begin{eqnarray}
&& S_{D0} = T_{D0} \int dt Tr( \Sigma_{m,n=0}^{9}
\partial_{t}X^{n}\partial_{t}X^{m}+...) \label{sfd8}
\end{eqnarray}
which is an approximation of following action:

\begin{eqnarray}
&& S_{D0} = T_{D0} \int dt Tr( \sqrt{1+
\partial_{t}X^{n}\partial_{t}X^{m}}+..) \label{sg8}
\end{eqnarray}

This action includes only derivatives respect to time which is only dimension of this brane.

Replacing two dimensional brackets in Lie-two-algebra by three dimensional brackets in Lie-three-algebra, we obtain the action of M0-branes\cite{h2,h3,h4,h5,h6,h7,h8,h9,h13,h14,h15,h16}:

\begin{eqnarray}
S_{M0} = T_{M0}\int dt Tr( \Sigma_{M,N,L=0}^{10}
\langle[X^{M},X^{N},X^{L}],[X^{M},X^{N},X^{L}]\rangle) \label{s9}
\end{eqnarray}

where $X^{M}=X^{M}_{\alpha}T^{\alpha}$ and

\begin{eqnarray}
 &&[T^{\alpha}, T^{\beta}, T^{\gamma}]= f^{\alpha \beta \gamma}_{\eta}T^{\eta} \nonumber \\&&\langle T^{\alpha}, T^{\beta} \rangle = h^{\alpha\beta} \nonumber \\&& [X^{M},X^{N},X^{L}]=[X^{M}_{\alpha}T^{\alpha},X^{N}_{\beta}T^{\beta},X^{L}_{\gamma}T^{\gamma}]\nonumber \\&&\langle X^{M},X^{M}\rangle = X^{M}_{\alpha}X^{M}_{\beta}\langle T^{\alpha}, T^{\beta} \rangle
\label{s10}
\end{eqnarray}

where  $X^{M}$(i=1,3,...10) are  scalar strings which are linked to M0-brane. Using $\langle[X^{j},X^{0},X^{i}],[X^{j},X^{0},X^{i}]\rangle=(X^{j}\varepsilon^{i0j}\partial_{t}X^{i})^{2}$ \cite{h2,h3,h4,h5}, the action transforms to :
 
\begin{eqnarray}
&& S_{M0} = T_{M0}\int dt Tr( \Sigma_{i,j=0}^{10}
(X^{j}\varepsilon^{i0j}\partial_{t}X^{i})^{2}+...) \label{s8}
\end{eqnarray}
which is an approximation of following action:

\begin{eqnarray}
&& S_{M0} = T_{M0} \int dt Tr( \sqrt{1+
(X^{j}\varepsilon^{i0j}\partial_{t}X^{i})(X^{k}\varepsilon^{i0j}\partial_{t}X^{l})}+..) \label{soo8}
\end{eqnarray}

This action also includes only derivatives respect to time which is only dimension of this brane in 11 dimensions.

  By substituting N-dimensional brackets instead of three dimensional brackets in action of  (\ref{s8}), we obtain the action of G0-brane in G-theory as:

 \begin{eqnarray}
 S_{G0} = T_{G0}\int dt Tr( \Sigma_{L_{1}=L_{2}..L_{N}=0}^{M}
 \langle[X^{L_{1}},X^{L_{2}},...X^{L_{N}}],[X^{L_{1}},X^{L_{2}},...X^{L_{N}}]\rangle) \label{P1}
 \end{eqnarray}
 
 where $X^{M}=X^{M}_{\alpha}T^{\alpha}$ and

 \begin{eqnarray}
  &&[T^{\alpha_{1}}, T^{\alpha_{2}}..T^{\alpha_{N}}]= f^{\alpha_{1}..\alpha_{N}}_{\alpha_{L}}T^{L} \nonumber \\&&\langle T^{\alpha}, T^{\beta} \rangle = h^{\alpha\beta} \nonumber \\&& [X^{L_{1}},X^{L_{2}},...X^{L_{N}}]=[X^{L_{1}}_{\alpha_{1}}T^{\alpha_{1}},X^{L_{2}}_{\alpha_{2}}T^{\alpha_{2}},...X^{L_{N}}_{\alpha_{N}}T^{\alpha_{N}}]\nonumber \\&&\langle X^{M},X^{M}\rangle = X^{M}_{\alpha}X^{M}_{\beta}\langle T^{\alpha}, T^{\beta} \rangle
 \label{P2}
 \end{eqnarray}
 
  Above action may transit to the action of M0-branes by puting N=3 and M=10 and also to the action of D0-brane for N=2 and M=9.  Now, the question arises that what is the origin of this brane in G-theory?  To answer this question, we suppose  that first, two energies with opposite signs are produced such as the sum over them be zero. Then, these energies  create 2M degrees of freedom which each two of them cause  to creation of new dimension. After that, M-N of degrees of freedom are hiden by compacting half of M-N dimensions on a circle to create Lie-N-algebra. During this process, the behaviour of some dimensions changes and they form times. Also, for second energy, the properties of more dimensions change which lead to the emergence of more time coordinates. Thus, physics of branes is completely different from anti-branes.
         
         Let us to begin with two oscillating energies which  are created from nothing and expanded in Mth dimension. We obtain:

           \begin{eqnarray}
           && E \equiv 0\equiv E_{1}+E_{2}\equiv 0\equiv N_{1}+N_{2}\equiv k((X^{M})^{2}-  (X^{M})^{2}) = k \int d^{2}x (\frac{\partial}{\partial x})^{2}((X^{M})^{2}-  (X^{M})^{2})
           \label{t1}
           \end{eqnarray}
         
         where $N_{1/2}$ are the number of degrees of freedom for first and second energies. These energies are oscillating, excited and produce M dimensions with 2M degrees of freedom. Each degree of freedom produces an integral and a derivative $\int d^{2}x (\frac{\partial}{\partial x})^{2}F=F$ such as the initial results don't change and the sum over energies becomes zero again. This can be shown clearly  by rewriting equation (\ref{t1})as follows:

             \begin{eqnarray}
             && E \equiv 0\equiv k \int d^{2M}x \varepsilon^{i_{1}i_{2}...i_{M}}\varepsilon^{i_{1}i_{2}...i_{M}} (\frac{\partial}{\partial x_{i_{1}}}\frac{\partial}{\partial x_{i_{2}}}..\frac{\partial}{\partial x_{i_{M-1}}})^{2}(X^{M})^{2}-\nonumber \\ &&  k \int d^{2M}x \varepsilon^{i_{1}i_{2}...i_{M}}\varepsilon^{i_{1}i_{2}...i_{M}} (\frac{\partial}{\partial x_{i_{1}}}\frac{\partial}{\partial x_{i_{2}}}..\frac{\partial}{\partial x_{i_{M-1}}})^{2}(X^{M})^{2}
             \label{t2}
             \end{eqnarray}
         
           where, we apply the difinition  $\varepsilon^{i_{1}i_{2}...i_{M}}\varepsilon^{i_{1}i_{2}...i_{M}}=-1$. We can substitute some brackets instead of derivatives and write \cite{h2,h3,h4,h5,h6,h13,h14,h15,h16}:

             \begin{eqnarray}
             && \frac{\partial}{\partial x_{i_{1}}}X^{M}=[ X^{i_{1}},X^{14}] \nonumber \\ && \frac{\partial}{\partial x_{i_{1}}}\frac{\partial}{\partial x_{i_{2}}}X^{M}=[ X^{i_{1}},X^{i_{2}},X^{M}]\nonumber \\ && (\frac{\partial}{\partial x_{i_{1}}}\frac{\partial}{\partial x_{i_{2}}}..\frac{\partial}{\partial x_{i_{M-1}}})(X^{M})=[ X^{i_{1}},X^{i_{2}},...,X^{i_{M-1}},X^{M}]\nonumber \\ &&\varepsilon^{i_{1}i_{2}...i_{M}}\varepsilon^{i_{1}i_{2}...i_{M}} (\frac{\partial}{\partial x_{i_{1}}}\frac{\partial}{\partial x_{i_{2}}}..\frac{\partial}{\partial x_{i_{M-1}}})^{2}(X^{M})^{2}=\nonumber \\ && \varepsilon^{i_{1}i_{2}...i_{M}}\varepsilon^{i'_{1}i'_{2}...i'_{M}} [(\frac{\partial}{\partial x_{i_{1}}}\frac{\partial}{\partial x_{i_{2}}}..\frac{\partial}{\partial x_{i_{M-1}}})(X^{M})][(\frac{\partial}{\partial x_{i'_{1}}}\frac{\partial}{\partial x_{i'_{2}}}..\frac{\partial}{\partial x_{i'_{M-1}}})(X^{M})]=\nonumber \\ && \langle [ X_{i_{1}},X_{i_{2}},...,X_{i_{M}}],[ X_{i_{1}},X_{i_{2}},
          ...,X_{i_{M}}] \rangle
             \label{t3}
             \end{eqnarray}
           
        Applying  the relations of equation (\ref{t3}) in equation (\ref{t2}), we get:

             \begin{eqnarray}
             && E \equiv 0\equiv E_{1}+E_{2}\equiv \nonumber \\ &&  E_{1}=  k \int d^{2M}x \langle [ X_{i_{1}},X_{i_{2}},...,X_{i_{M}}],[ X_{i_{1}},X_{i_{2}},...,X_{i_{M}}] \rangle \nonumber \\ &&   E_{2}=-k \int d^{2M}x \langle [ X_{i_{1}},X_{i_{2}},...,X_{i_{M}}],[ X_{i_{1}},X_{i_{2}},...,X_{i_{M}}] \rangle
             \label{t4}
             \end{eqnarray}  
             
           The shape of above energies is similar to the shape of action of Gp-branes, however  their algebra has M dimensional bracket, while G0-brane has N dimensional bracket and for producing this algebra, we should remove M-N of degrees of freedom by compacting. To achieve this aim, we apply the mechanism in  \cite{h16} and replace $X_{i_{n=1,3,5..M-N}}=i T^{i_{n}}\frac{R}{l_{P}^{1/2}}$   where $l_{P}$ is the Planck length and ($(T^{i_{n}})^{2}=1$). We derive the below action for first energy:

               \begin{eqnarray}
               && E_{1}\equiv k \int d^{2M}x \langle [ X_{i_{1}},X_{i_{2}},...,X_{i_{M}}],
               [X_{i_{1}},X_{i_{2}},...,X_{i_{M}}] \rangle=
               \nonumber\\&& k \int d^{2M}x  \varepsilon^{i_{1}i_{2}...i_{M}}\varepsilon^{i'_{1}i'_{2}...i'_{M}}X_{i_{1}}X_{i_{2}}...X_{i_{M}}X_{i'_{1}}X_{i'_{2}}...X_{i'_{M}} = \nonumber \\ && (i)^{2(M-N)}k \int d^{N}x (\frac{R^{M-N}}{l_{P}^{(M-N)/2}}) \varepsilon^{j_{1}...j_{N}}\varepsilon^{j'_{1}...j'_{N}}X_{j_{1}}...X_{j_{N}}X_{j'_{1}}...
               X_{j'_{N}}=\nonumber\\ && (i)^{2(M-N)}k \int d^{N}x (\frac{R^{M-N}}{l_{P}^{(M-N)/2}})\langle [ X_{j_{1}},X_{j_{2}},...,X_{j_{N}}],[ X_{j_{1}},X_{j_{2}},...,X_{j_{N}}]\rangle=\nonumber\\ && k \int d^{N}x (\frac{R^{M-N}}{l_{P}^{(M-N)/2}}) \langle [i X_{j_{1}},iX_{j_{2}},...,iX_{j_{M-N}}..,X_{j_{N}}],[ iX_{j_{1}},X_{j_{2}},..,...,iX_{j_{M-N}}.,X_{j_{N}}]\rangle
              \label{t5}
               \end{eqnarray}  
             
              where we have defined $ \varepsilon^{1i_{2}...i_{M}}\varepsilon^{1i'_{2}...i'_{M}}=(-i)^{N-M}\varepsilon^{j_{1}...j_{N}}\varepsilon^{j'_{1}...j'_{N}}$. Some scalar strings take  one extra (i) and  expands in time directions. Obviously, in G-theory, there exists M-N time coordinates where M is dimension of world and N is dimension of algebra. The reason that we only observe one dimension is  our living in four dimensional universe and observing three dimensional brackets of M-theory. For this reason, for us, M=4 and n=3 and thus we have only one time dimension. For second energy which  it's sign is opposite to first energy, we have some extra time dimensions:

                   \begin{eqnarray}
                   && E_{2}=-E_{1}=(i)^{2}E_{1}= \nonumber\\ && k \int d^{N}x (\frac{R^{M-N}}{l_{P}^{(M-N)/2}}) \langle [i X_{j_{1}},iX_{j_{2}},...,iX_{j_{M-N+1}}..,X_{j_{N}}],[ iX_{j_{1}},X_{j_{2}},..,...,iX_{j_{M-N+1}}.,X_{j_{N}}]\rangle  \label{t6}
                   \end{eqnarray}  
                   
             Thus, properties of anti-branes branes which are created by this energy is different and we have more time dimensions. For example, in one four dimensional anti-universe, there exists  two time coordinates and all things are changed. In our universe, length of one object can be defined by $l^{2}=-t^{2}+x_{1}^{2}+x_{2}^{2}+x_{3}^{2}$ where t is time and $x_{1}$ are coordinates of space. While,  in anti-universe, length is obtained by $\tilde{l}^{2}=-t_{1}^{2}-x_{1}^{2}+x_{2}^{2}+x_{3}^{2}$. Also, energy and momentums which is related to the mass with this equation for our universe ($m^{2}=-E^{2}+P_{1}^{2}+P_{2}^{2}+P_{3}^{2}$), has this relation ($m^{2}=-E^{2}-P_{1}^{2}+P_{2}^{2}+P_{3}^{2}$) for anti-universe.

  Until now, we have considered only symmetrical compacting, however maybe, during this process, the symmetry of system is broken and only  upper or lower part of one dimension is compactified and fermions are emerged($X \rightarrow \psi^{U}\psi^{L}$ ) \cite{h3,h4}. To include non-symmetrical compacting, we use the mechanism in \cite{h16}, and compactify Mth dimension of branes on a circle with radius R by choosing  $<X^{M}>=i\frac{R}{l_{p}^{1/2}}T^{M}$ for boson and $<\psi^{L,M}>=i\frac{R^{1/2}}{l_{p}^{1/4}}T^{L,M}$ for fermions in action of (\ref{t5}). Generators of algebra for bosons and fermions have a direct relation with each other (($(T^{i_{n}})^{2}=1$) and $T^{L,M}T^{R,M}=T^{M}$). We get: 
   
                  \begin{eqnarray}
                  && E_{1}\equiv  k \int d^{N}x (\frac{R^{M-N}}{l_{P}^{(M-N)/2}})\Big( \langle [i X_{j_{1}},iX_{j_{2}},...,iX_{j_{M-N}}..,X_{j_{N}}],[ iX_{j_{1}},X_{j_{2}},..,...,iX_{j_{M-N}}.,X_{j_{N}}]\rangle   \nonumber \\
                                  && 
                            - i(\frac{R^{M-N}}{l_{P}^{(M-N)/2}})\langle [i X_{j_{1}},iX_{j_{2}},...,iX_{j_{M-N}}..T_{j_{m'}}.,\psi_{R,j_{N}}], [i X_{j_{1}},iX_{j_{2}},...,iX_{j_{M-N}}..T_{j_{m'}}.,\psi_{R,j_{N}}]\rangle ) )\Big)
                 \label{P6}
                  \end{eqnarray}  
     
  By choosing $\gamma_{j_{m}}=T_{j_{m}}\frac{R^{2}}{l_{p}}$ where $\gamma^{L_{1}}$'s are the pauli matrices in M dimensions, we obtain the  action of initial energy as follows:

                    \begin{eqnarray}
                    && E_{1}\equiv  k \int d^{N}x  (\frac{R^{M-N}}{l_{P}^{(M-N)/2}})\Big(\langle [i X_{j_{1}},iX_{j_{2}},...,iX_{j_{M-N}}..,X_{j_{N}}],[ iX_{j_{1}},X_{j_{2}},..,...,iX_{j_{M-N}}.,X_{j_{N}}]\rangle  \nonumber \\
                                    && 
                              -  i\langle [i X_{j_{1}},iX_{j_{2}},...,iX_{j_{M-N}}..\gamma_{j_{m'}}.,\psi_{R,j_{N}}], [i X_{j_{1}},iX_{j_{2}},...,iX_{j_{M-N}}..\gamma_{j_{m'}}.,\psi_{R,j_{N}}]\rangle )\Big)
                   \label{P7}
                    \end{eqnarray}  
  
  This action includes  both fermionic and bosonic degrees of freedom and supersymmetry emerges. If we assume that for an object with one time dimension, all scalars depend only on one time coordinates and ($R=l_{P}^{1/2}$) \cite{h16} , we achieve to action of G0-branes:

                     \begin{eqnarray}
                     && S_{G0}\equiv  k V_{N-1} \int dt \Big(\langle [i X_{j_{1}},iX_{j_{2}},...,iX_{j_{M-N}}..,X_{j_{N}}],[ iX_{j_{1}},X_{j_{2}},..,...,iX_{j_{M-N}}.,X_{j_{N}}]\rangle   \nonumber \\
                                     && 
                               -  i\langle [i X_{j_{1}},iX_{j_{2}},...,iX_{j_{M-N}}..\gamma_{j_{m'}}.,\psi_{R,j_{N}}], [i X_{j_{1}},iX_{j_{2}},...,iX_{j_{M-N}}..\gamma_{j_{m'}}.,\psi_{R,j_{N}}]\rangle )\Big)
                    \label{PP7}
                     \end{eqnarray}
                     
    where $V_{N-1}$ is the volum of space which is formed by remaining coordinates.  Also, by adding one negative sign and using equation (\ref{t6}), we obtain the action of anti-G0-brane:  
                  
                                          \begin{eqnarray}
                                          && S_{Anti-G0}\equiv  k V_{N-1} \int dt  \Big(\langle [i X_{j_{1}},iX_{j_{2}},...,iX_{j_{M-N+1}}..,X_{j_{N}}],[ iX_{j_{1}},X_{j_{2}},..,...,iX_{j_{M-N+1}}.,X_{j_{N}}]\rangle   \nonumber \\
                                                          && 
                                                    -  i\langle [i X_{j_{1}},iX_{j_{2}},...,iX_{j_{M-N+1}}..\gamma_{j_{m'}}.,\psi_{R,j_{N}}], [i X_{j_{1}},iX_{j_{2}},...,iX_{j_{M-N+1}}..\gamma_{j_{m'}}.,\psi_{R,j_{N}}]\rangle )\Big)
                                         \label{PPP7}
                                          \end{eqnarray}

  These actions for G0-brane and G0-anti-brane contain both bosonic and fermionic fields which are created as due to symmetrical or non-symmetrical compacting of dimensions. Also, time which is a puzzle in cosmology is produced by  compacting. Thus, we conclude that all fields have the same origin and  begin from nothing. Then, by different compactification, different shapes of matters emerge.  
  
  By substituting N-dimensional brackets in equations (\ref{PP7}) and (\ref{PPP7}) instead of two dimensional brackets  and increasing dimensions from 10 to M in action (\ref{s7}), we can obtain the action of Gp-brane and anti-Gp-brane:

               \begin{eqnarray}
               &&S_{Gp} = -(T_{G0})^{p} \int dt \sum_{n=1}^{p}\beta_{n}\Big(
               \delta^{a_{1},a_{2}...a_{n}}_{b_{1}b_{2}....b_{n}}L^{b_{1}}_{a_{1}}...L^{b_{n}}_{a_{n}}\Big)^{1/2}\nonumber\\&&
               (L)^{a_{n}}_{b_{n}}= \delta^{a_{n}}_{b_{n}}Tr\Big( \Sigma_{L=0}^{N} \Sigma_{H=0}^{N-L}\Sigma_{a_{1}..a_{L}=0}^{p}\Sigma_{j_{1}..j_{H}=p+1}^{M}(\nonumber \\
                                 && i^{2(p-N)}\langle[X^{j_{1}},..X^{j_{H-1}},X^{a_{1}},..X^{a_{L}},X^{j_{H}}],\langle[X^{j_{1}},..X^{j_{H-1}},X^{a_{1}},..X^{a_{L}},X^{j_{H}}]\rangle) + \nonumber \\
                      && i^{2(p-N)}\Sigma_{L=0}^{N} \Sigma_{H=0}^{N-L}\Sigma_{a_{1}..a_{L}=0}^{p}\Sigma_{j_{1}..j_{H}=p+1}^{M}
                           (\langle[X^{j_{1}},..X^{j_{H}},X^{a_{1}},..X^{a_{L}}],[X^{j_{1}},..X^{j_{H}},X^{a_{1}},..X^{a_{L}}]\rangle)- \nonumber \\
                                     &&  i^{2(p-N)+1}\langle[\gamma^{j_{1}},..X^{j_{H-1}},X^{a_{1}},..X^{a_{L}},\psi^{R,j_{H}}],\langle[X^{j_{1}},..X^{j_{H-1}},X^{a_{1}},..X^{a_{L}},\psi^{R,j_{H}}]\rangle) - \nonumber \\
                                               && i^{2(p-N)+1}\langle[\gamma^{j_{1}},..X^{j_{H}},X^{a_{1}},..X^{a_{L}-1},\psi^{R,a_{L}}],\langle[X^{j_{1}},..X^{j_{H}},X^{a_{1}},..X^{a_{L}-1},\psi^{R,a_{L}}]\rangle)\Big) \label{P88}
               \end{eqnarray}

                           \begin{eqnarray}
                           &&S_{Anti-Gp} = -(T_{Anti-G0})^{p} \int dt \sum_{n=1}^{p}\beta_{n}\Big(
                           \delta^{a_{1},a_{2}...a_{n}}_{b_{1}b_{2}....b_{n}}L^{b_{1}}_{a_{1}}...L^{b_{n}}_{a_{n}}\Big)^{1/2}\nonumber\\&&
                           (L)^{a_{n}}_{b_{n}}= \delta^{a_{n}}_{b_{n}}Tr\Big( \Sigma_{L=0}^{N} \Sigma_{H=0}^{N-L}\Sigma_{a_{1}..a_{L}=0}^{p}\Sigma_{j_{1}..j_{H}=p+1}^{M}(\nonumber \\
                                             && i^{2(p-N+1)}\langle[X^{j_{1}},..X^{j_{H-1}},X^{a_{1}},..X^{a_{L}},X^{j_{H}}],\langle[X^{j_{1}},..X^{j_{H-1}},X^{a_{1}},..X^{a_{L}},X^{j_{H}}]\rangle) + \nonumber \\
                                  && i^{2(p-N+1)}\Sigma_{L=0}^{N} \Sigma_{H=0}^{N-L}\Sigma_{a_{1}..a_{L}=0}^{p}\Sigma_{j_{1}..j_{H}=p+1}^{M}
                                       (\langle[X^{j_{1}},..X^{j_{H}},X^{a_{1}},..X^{a_{L}}],[X^{j_{1}},..X^{j_{H}},X^{a_{1}},..X^{a_{L}}]\rangle)- \nonumber \\
                                                 &&  i^{2(p-N+1)+1}\langle[\gamma^{j_{1}},..X^{j_{H-1}},X^{a_{1}},..X^{a_{L}},\psi^{R,j_{H}}],\langle[X^{j_{1}},..X^{j_{H-1}},X^{a_{1}},..X^{a_{L}},\psi^{R,j_{H}}]\rangle) - \nonumber \\
                                                           && i^{2(p-N+1)+1}\langle[\gamma^{j_{1}},..X^{j_{H}},X^{a_{1}},..X^{a_{L}-1},\psi^{R,a_{L}}],\langle[X^{j_{1}},..X^{j_{H}},X^{a_{1}},..X^{a_{L}-1},\psi^{R,a_{L}}]\rangle)\Big) \label{PP8}
                           \end{eqnarray}

 To write actions in terms of gauge fields and derivatives respect to fields, we have to use of some laws. Previously, some rules have been discussed in \cite{h2,h3,h4,h5,h15} for Lie-three-algebra. These laws  have some origins in topology inadditional to mathematical methodes and M-theory. In these brackets, "$X$" is the scalar string. When, in one bracket, there is only one indice related to brane and before other indices which are related to transverse directions respect to brane, it means that one string has been glued from one end to brane and we can write one derivate of string respect to coordinates of brane. When, one bracket includes two indices of branes after indices of transverse directions respect to brane, two ends of string has been attached and some gauge fields like photons are appeared. The place of indices is also important. For example, if two indices of branes be before one indice of transverse direction, two derivatives are appeared for strings. Also, if two or more indices of branes be as the last indices of bracket, tensor fields are emerged that their rank is equal to number of indices.  Extending the rules in equation (\ref{r3}) for M-theory to N-dimensional brackets in G-theory, we can obtain following laws \cite{h2,h3,h4,h5,h10,h11,h12,h13,h14,h15,h16}:

  \begin{eqnarray}
  && \Sigma_{L=0}^{N} \Sigma_{H=0}^{N-L}\Sigma_{a_{1}..a_{L}=0}^{p}\Sigma_{j_{1}..j_{H}=p+1}^{M} \langle[X^{j_{1}},..X^{j_{H-1}},X^{a_{1}},..X^{a_{L}},X^{j_{H}}],\langle[X^{j_{1}},..X^{j_{H-1}},X^{a_{1}},..X^{a_{L}},X^{j_{H}}]\rangle=\nonumber \\
       &&
   \frac{1}{2}\Sigma_{L=0}^{N} \Sigma_{H=0}^{N-L}\Sigma_{a_{1}..a_{L}=0}^{p}\Sigma_{j_{1}..j_{H}=p+1}^{M}(X^{j_{1}}..X^{j_{H-1}})^{2}\langle \partial_{a_{1}}..\partial_{a_{L}}X^{i},\partial_{a_{1}}..\partial_{a_{L}}X^{i}\rangle\nonumber \\
     &&\nonumber \\
       &&\nonumber \\
  && \Sigma_{L=0}^{N} \Sigma_{H=0}^{N-L}\Sigma_{a_{1}..a_{L}=0}^{p}\Sigma_{j_{1}..j_{H}=p+1}^{M}
     \langle[X^{j_{1}},..X^{j_{H}},X^{a_{1}},..X^{a_{L}}],[X^{j_{1}},..X^{j_{H}},X^{a_{1}},..X^{a_{L}}]\rangle=\nonumber \\
         &&
        \Sigma_{L=0}^{N} \Sigma_{H=0}^{N-L}\Sigma_{a_{1}..a_{L}=0}^{p}\Sigma_{j_{1}..j_{H}=p+1}^{M}\frac{\lambda^{2}}{1.2...N}(X^{j_{1}}..X^{j_{H}})^{2}\langle F^{a_{1}..a_{L}}, F^{a_{1}..a_{L}}\rangle\nonumber \\
  &&\nonumber \\
    &&\nonumber \\
      &&\nonumber \\
        &&F_{a_{1}..a_{n}}=\partial_{[a_{1}} A_{a_{2}..a_{n}]}=\partial_{a_{1}} A_{a_{2}..a_{n}}-\partial_{a_{2}} A_{a_{1}..a_{n}}+..\nonumber \\
          &&\nonumber \\
            &&\nonumber \\
  &&\Sigma_{m}\rightarrow \frac{1}{(2\pi)^{p}}\int d^{p+1}\sigma \Sigma_{m-p-1}
  i,j=p+1,..,M\quad a,b=0,1,...p\quad m,n=0,..,M~~  
 \nonumber \\
      && \nonumber \\
        && \nonumber \\
          && \Sigma_{L=0}^{N} \Sigma_{H=0}^{N-L}\Sigma_{a_{1}..a_{L}=0}^{p}\Sigma_{j_{1}..j_{H}=p+1}^{M} \langle[\gamma^{j_{1}},..X^{j_{H-1}},X^{a_{1}},..X^{a_{L}},\psi^{j_{H}}],\langle[X^{j_{1}},..X^{j_{H-1}},X^{a_{1}},..X^{a_{L}},\psi^{j_{H}}]\rangle=\nonumber \\
         &&
     \frac{1}{2}i\Sigma_{L=0}^{N} \Sigma_{H=0}^{N-L}\Sigma_{a_{1}..a_{L}=0}^{p}\Sigma_{j_{1}..j_{H}=p+1}^{M}(X^{j_{1}}..X^{j_{H-1}})^{2}\gamma^{a_{L-1}}\langle \partial_{a_{1}}..\partial_{a_{L-1}}\psi^{i},\partial_{a_{1}}..\partial_{a_{L}}\psi^{i}\rangle\nonumber \\
       &&\nonumber \\
         &&\nonumber \\
    &&  \Sigma_{L=0}^{N} \Sigma_{H=0}^{N-L}\Sigma_{a_{1}..a_{L}=0}^{p}\Sigma_{j_{1}..j_{H}=p+1}^{M} \langle[\gamma^{j_{1}},..X^{j_{H-1}},X^{a_{1}},..X^{a_{L}-1},\psi^{a_{L}}],\langle[X^{j_{1}},..X^{j_{H}},X^{a_{1}},..X^{a_{L}-1},\psi^{a_{L}}]\rangle=\nonumber \\
           &&
          i\Sigma_{L=0}^{N} \Sigma_{H=0}^{N-L}\Sigma_{a_{1}..a_{L}=0}^{p}\Sigma_{j_{1}..j_{H}=p+1}^{M}\frac{\lambda^{2}}{1.2...N}(X^{j_{1}}..X^{j_{H}})^{2}\gamma^{a_{L-1}}\langle \bar{F}^{a_{1}..a_{L-1}}, \bar{F}^{a_{1}..a_{L}}\rangle\nonumber \\
    &&\nonumber \\
      &&\nonumber \\
        &&\nonumber \\
          &&\bar{F}_{a_{1}..a_{n}}=\partial_{[a_{1}} \bar{A}_{a_{2}..a_{n}]}=\partial_{a_{1}} \bar{A}_{a_{2}..a_{n}}-\partial_{a_{2}} \bar{A}_{a_{1}..a_{n}}+..\nonumber \\
            &&\nonumber \\
              &&\nonumber \\
    &&\Sigma_{m}\rightarrow \frac{1}{(2\pi)^{p}}\int d^{p+1}\sigma \Sigma_{m-p-1}
    i,j=p+1,..,M\quad a,b=0,1,...p\quad m,n=0,..,M~~
    \label{P9}
    \end{eqnarray}
       
    Here $\bar{A}_{a_{2}..a_{n}}$ are fermionic superpartners of gauge bosons $A_{a_{2}..a_{n}}$ and $\psi$  are the fermionic superpartner of scalar strings $X$.  
     Using rules of equation (\ref{P9}) in action (\ref{PP8} and \ref{P88} ), we derive the following action for Gp-branes:

          \begin{eqnarray}
          &&S_{Gp} = -(T_{Gp}) \int dt \sum_{n=1}^{p}\beta_{n}\Big(
          \delta^{a_{1},a_{2}...a_{n}}_{b_{1}b_{2}....b_{n}}L^{b_{1}}_{a_{1}}...L^{b_{n}}_{a_{n}}\Big)^{1/2}\nonumber\\&&
          (L)^{a_{n}}_{b_{n}}= \delta^{a_{n}}_{b_{n}}Tr\Big( 
             \frac{1}{2}i^{2(p-N)}\Sigma_{L=0}^{N} \Sigma_{H=0}^{N-L}\Sigma_{a_{1}..a_{L}=0}^{p}\Sigma_{j_{1}..j_{H}=p+1}^{M} (X^{j_{1}}..X^{j_{H-1}})^{2}\langle \partial_{a_{1}}..\partial_{a_{L}}X^{i},\partial_{a_{1}}..\partial_{a_{L}}X^{i}\rangle + \nonumber \\
                 && 
                        i^{2(p-N)} \Sigma_{L=0}^{N} \Sigma_{H=0}^{N-L}\Sigma_{a_{1}..a_{L}=0}^{p}\Sigma_{j_{1}..j_{H}=p+1}^{M}\frac{\lambda^{2}}{1.2...N} (X^{j_{1}}..X^{j_{H}})^{2}\langle F^{a_{1}..a_{L}}, F^{a_{1}..a_{L}}\rangle  -\nonumber\\ &&
                           \frac{1}{2}i^{2(p-N)+1}\Sigma_{L=0}^{N} \Sigma_{H=0}^{N-L}\Sigma_{a_{1}..a_{L}=0}^{p}\Sigma_{j_{1}..j_{H}=p+1}^{M} (X^{j_{1}}..X^{j_{H-1}})^{2}\gamma^{a_{L-1}}\langle \partial_{a_{1}}..\partial_{a_{L-1}}\psi^{i},\partial_{a_{1}}..\partial_{a_{L}}\psi^{i}\rangle -\nonumber \\
                               &&  i^{2(p-N)+1}\Sigma_{L=0}^{N} \Sigma_{H=0}^{N-L}\Sigma_{a_{1}..a_{L}=0}^{p}\Sigma_{j_{1}..j_{H}=p+1}^{M}\frac{\lambda^{2}}{1.2...N}\gamma^{a_{L-1}}(X^{j_{1}}..X^{j_{H}})^{2} \langle \bar{F}^{a_{1}..a_{L-1}}, \bar{F}^{a_{1}..a_{L}}\rangle \Big) \label{P10}
          \end{eqnarray}

                 \begin{eqnarray}
                 &&S_{Anti-Gp} = -(T_{Anti-Gp}) \int dt \sum_{n=1}^{p}\beta_{n}\Big(
                 \delta^{a_{1},a_{2}...a_{n}}_{b_{1}b_{2}....b_{n}}L^{b_{1}}_{a_{1}}...L^{b_{n}}_{a_{n}}\Big)^{1/2}\nonumber\\&&
                 (L)^{a_{n}}_{b_{n}}= \delta^{a_{n}}_{b_{n}}Tr\Big( 
                    \frac{1}{2}i^{2(p-N+1)}\Sigma_{L=0}^{N} \Sigma_{H=0}^{N-L}\Sigma_{a_{1}..a_{L}=0}^{p}\Sigma_{j_{1}..j_{H}=p+1}^{M} (X^{j_{1}}..X^{j_{H-1}})^{2}\langle \partial_{a_{1}}..\partial_{a_{L}}X^{i},\partial_{a_{1}}..\partial_{a_{L}}X^{i}\rangle + \nonumber \\
                        && 
                               i^{2(p-N+1)} \Sigma_{L=0}^{N} \Sigma_{H=0}^{N-L}\Sigma_{a_{1}..a_{L}=0}^{p}\Sigma_{j_{1}..j_{H}=p+1}^{M}\frac{\lambda^{2}}{1.2...N} (X^{j_{1}}..X^{j_{H}})^{2}\langle \hat{F}^{a_{1}..a_{L}}, \hat{F}^{a_{1}..a_{L}}\rangle  -\nonumber\\ &&
                                  \frac{1}{2}i^{2(p-N+1)+1}\Sigma_{L=0}^{N} \Sigma_{H=0}^{N-L}\Sigma_{a_{1}..a_{L}=0}^{p}\Sigma_{j_{1}..j_{H}=p+1}^{M} (X^{j_{1}}..X^{j_{H-1}})^{2}\gamma^{a_{L-1}}\langle \partial_{a_{1}}..\partial_{a_{L-1}}\psi^{i},\partial_{a_{1}}..\partial_{a_{L}}\psi^{i}\rangle -\nonumber \\
                                      &&  i^{2(p-N+1)+1}\Sigma_{L=0}^{N} \Sigma_{H=0}^{N-L}\Sigma_{a_{1}..a_{L}=0}^{p}\Sigma_{j_{1}..j_{H}=p+1}^{M}\frac{\lambda^{2}}{1.2...N}\gamma^{a_{L-1}}(X^{j_{1}}..X^{j_{H}})^{2} \langle \hat{\bar{F}}^{a_{1}..a_{L-1}}, \hat{\bar{F}}^{a_{1}..a_{L}}\rangle \Big) \label{PP10}
                 \end{eqnarray}
              
   These actions transit to action of Dp-brane (\ref{s7}) for N=2 and M=9 and action of Mp-brane  for N=3 and M=10 \cite{h2,h3,h4,h5,h6,h7,h8,h9,h10,h11,h12,h13,h14,h15,h16} . Also, it is clear that for each scalar, there is a fermion and for each bosonic gauge field with each rank, there exists a fermionic gauge field with the same rank. This means that these actions are supersymmetric and the number of degrees of freedom for both fermions and bosons are the same.  We can show that above actions are invariant under following transformations:
      
     \begin{eqnarray}
     && \delta X^{i}= \psi^{i} \nonumber\\&& \delta \partial_{a_{j}}\psi^{i} =i \gamma^{a_{j}}\partial_{a_{j}}\partial_{a_{j}+1}X^{i} \nonumber\\&& \delta F^{a_{1}..a_{L}}= \bar{F}^{a_{1}..a_{L}} \nonumber\\&& \delta \bar{F}^{a_{1}..a_{j}}= i\gamma^{a_{j}}F^{a_{1}..a_{j+1}} \nonumber\\&& \delta \hat{F}^{a_{1}..a_{L}}= \hat{\bar{F}}^{a_{1}..a_{L}}\nonumber\\&& \delta \hat{\bar{F}}^{a_{1}..a_{j}}= i \gamma^{a_{j}}\hat{F}^{a_{1}..a_{j+1}}\nonumber\\&&\gamma^{a_{j}}\gamma^{a_{j}+1}=0\nonumber\\&& \gamma^{a_{j}}\gamma^{a_{j}}=1\nonumber\\&& \psi^{i,L}X^{i}=\psi^{i,L}\psi^{i,L}\psi^{i,R}=0 \label{sp9}
     \end{eqnarray}  
      
    These transformations can be obtained from the laws of compactification in \cite{h16} and usual supersymmetric transformations which have been considered in many papers. In fact, by these transformations, bosons transverse to fermions and fermions changes to bosons.  With these transformations, we have: 
      \begin{eqnarray}
                && \delta S_{Gp} = -(T_{Gp}) \int dt \sum_{n=1}^{p}\beta_{n}\delta L^{b_{j}}_{a_{j}}(L^{b_{j}}_{a_{j}})^{-1/2}\Big(
                \delta^{a_{1},a_{2}...a_{n}}_{b_{1}b_{2}....b_{n}}L^{b_{1}}_{a_{1}}...L^{b_{n}}_{a_{n}}\Big)^{1/2}=0\nonumber\\&&
                \delta (L)^{a_{n}}_{b_{n}}= \delta^{a_{n}}_{b_{n}}Tr\Big( 
                   \frac{1}{2}i^{2(p-N)}\Sigma_{L=0}^{N} \Sigma_{H=0}^{N-L}\Sigma_{a_{1}..a_{L}=0}^{p}\Sigma_{j_{1}..j_{H}=p+1}^{M}\times\nonumber \\
                       && (X^{j_{1}}..X^{j_{H-1}})^{2}\langle \partial_{a_{1}}..\partial_{a_{L}}\psi^{i},\partial_{a_{1}}..\partial_{a_{L}}X^{i}\rangle + \nonumber \\
                       && 
                              i^{2(p-N)} \Sigma_{L=0}^{N} \Sigma_{H=0}^{N-L}\Sigma_{a_{1}..a_{L}=0}^{p}\Sigma_{j_{1}..j_{H}=p+1}^{M}\times\nonumber \\
                                  &&\frac{\lambda^{2}}{1.2...N} (X^{j_{1}}..X^{j_{H}})^{2}\langle \bar{F}^{a_{1}..a_{L}}, F^{a_{1}..a_{L}}\rangle  -\nonumber\\ &&
                                 \frac{1}{2}i^{2(p-N)}\Sigma_{L=0}^{N} \Sigma_{H=0}^{N-L}\Sigma_{a_{1}..a_{L}=0}^{p}\Sigma_{j_{1}..j_{H}=p+1}^{M}\times\nonumber \\
                                     && (X^{j_{1}}..X^{j_{H-1}})^{2}\langle \partial_{a_{1}}..\partial_{a_{L}}\psi^{i},\partial_{a_{1}}..\partial_{a_{L}}X^{i}\rangle -\nonumber \\
                                     &&  i^{2(p-N)}\Sigma_{L=0}^{N} \Sigma_{H=0}^{N-L}\Sigma_{a_{1}..a_{L}=0}^{p}\Sigma_{j_{1}..j_{H}=p+1}^{M}\times\nonumber \\
                                         &&\frac{\lambda^{2}}{1.2...N}(X^{j_{1}}..X^{j_{H}})^{2} \langle \bar{F}^{a_{1}..a_{L}}, F^{a_{1}..a_{L}}\rangle \Big)=0 \label{Pvb10}
                \end{eqnarray}
             
     \begin{eqnarray}
                      && \delta S_{Anti-Gp} = \nonumber\\&& -(T_{Anti-Gp}) \int dt \sum_{n=1}^{p}\beta_{n}\delta L^{b_{j}}_{a_{j}}(L^{b_{j}}_{a_{j}})^{-1/2}\Big(
                      \delta^{a_{1},a_{2}...a_{n}}_{b_{1}b_{2}....b_{n}}L^{b_{1}}_{a_{1}}...L^{b_{n}}_{a_{n}}\Big)^{1/2}=0\nonumber\\&&
                      \delta (L)^{a_{n}}_{b_{n}}= \delta^{a_{n}}_{b_{n}}Tr\Big( 
                         \frac{1}{2}i^{2(p-N+1)}\Sigma_{L=0}^{N} \Sigma_{H=0}^{N-L}\Sigma_{a_{1}..a_{L}=0}^{p}\Sigma_{j_{1}..j_{H}=p+1}^{M}\times \nonumber\\&& (X^{j_{1}}..X^{j_{H-1}})^{2}\langle \partial_{a_{1}}..\partial_{a_{L}}\psi^{i},\partial_{a_{1}}..\partial_{a_{L}}X^{i}\rangle + \nonumber \\
                             && 
                                    i^{2(p-N+1)} \Sigma_{L=0}^{N} \Sigma_{H=0}^{N-L}\Sigma_{a_{1}..a_{L}=0}^{p}\Sigma_{j_{1}..j_{H}=p+1}^{M}\times \nonumber\\&& \frac{\lambda^{2}}{1.2...N} (X^{j_{1}}..X^{j_{H}})^{2}\langle \hat{\bar{F}}^{a_{1}..a_{L}}, \hat{F}^{a_{1}..a_{L}}\rangle  -\nonumber\\ &&
                                       \frac{1}{2}i^{2(p-N+1)}\Sigma_{L=0}^{N} \Sigma_{H=0}^{N-L}\Sigma_{a_{1}..a_{L}=0}^{p}\Sigma_{j_{1}..j_{H}=p+1}^{M}\times \nonumber\\&&  (X^{j_{1}}..X^{j_{H-1}})^{2}\langle \partial_{a_{1}}..\partial_{a_{L}}\psi^{i},\partial_{a_{1}}..\partial_{a_{L}}X^{i}\rangle -\nonumber \\
                                           &&  i^{2(p-N+1)}\Sigma_{L=0}^{N} \Sigma_{H=0}^{N-L}\Sigma_{a_{1}..a_{L}=0}^{p}\Sigma_{j_{1}..j_{H}=p+1}^{M}\times \nonumber\\&& \frac{\lambda^{2}}{1.2...N}(X^{j_{1}}..X^{j_{H}})^{2} \langle \hat{\bar{F}}^{a_{1}..a_{L}}, \hat{F}^{a_{1}..a_{L}}\rangle \Big)=0 \label{PBVP10}
                      \end{eqnarray} 
      Thus, actions in this theory are invariant under supersymmetric transformations completely. This means that first, two positive and negative energies are emerged. Then, these energies produce two types of branes with bosonic fields on them. After that by compacting branes, superpartner of bosonic fields are emerged and supersymmetry is produced.            
   
      Now, we like to produce the supergravity in this new G-theory. For this reason, we assume that  $A^{ab}$ has the role of graviton and $\bar{A}^{ab}$ plays the role of gravitino and other higher dimensional fields have the below relations with these particles:

            \begin{eqnarray}
            && A_{a'b'}\rightarrow g_{a'b'} \nonumber \\
            && F_{a'b'c'}=\partial_{[a'} g_{b'c']}=\partial_{a'} g_{b'c'}-\partial_{c'} g_{a'b'}+\partial_{b'} g_{c'a'}\rightarrow  \Gamma _{a'b'c'
                 }\label{t28}
            \end{eqnarray}

            \begin{eqnarray}
            && A_{a'b'c'}\rightarrow F_{a'b'c'}\rightarrow\Gamma_{a'b'c'} \nonumber \\
                 && \tilde{R}_{ca'b'c'}\approx \partial_{[c} A_{a'b'c']} + \langle F_{\lambda ca'},F_{ b'c'}^{\lambda}\rangle \approx \partial_{[c}\Gamma _{a'b'c']}+ \Gamma_{\lambda ca'}\Gamma_{ b'c'}^{\lambda}-\Gamma_{\lambda cb'}\Gamma_{ a'c'}^{\lambda}\label{t29}
            \end{eqnarray}

                \begin{eqnarray}
                && A_{a'b'c'c}\rightarrow F_{a'b'c'c}\rightarrow \tilde{R}_{a'b'c'c} \nonumber \\
                          &&  F_{a_{1},a_{2},..a_{n}}=\partial_{[a_{5}}.. \partial_{a_{n}}\tilde{R}_{a_{1},a_{2},a_{3},a_{4}
                          ]}\label{t30}
                \end{eqnarray}

                            \begin{eqnarray}
                            && \bar{A}_{a'b'}\rightarrow \bar{g}_{a'b'} \nonumber \\
                            && \bar{F}_{a'b'c'}=\partial_{[a'} g_{b'c']}=\partial_{a'} g_{b'c'}-\partial_{c'} g_{a'b'}+\partial_{b'} g_{c'a'}\rightarrow  \Gamma _{a'b'c'
                                 }\label{tt28}
                            \end{eqnarray}

                            \begin{eqnarray}
                            && \bar{A}_{a'b'c'}\rightarrow \bar{F}_{a'b'c'}\rightarrow\Gamma_{a'b'c'} \nonumber \\
                                 && \bar{R}_{ca'b'c'}\approx \partial_{[c} A_{a'b'c']} + \langle \bar{F}_{\lambda ca'},\bar{F}_{ b'c'}^{\lambda}\rangle \approx \partial_{[c}\Gamma _{a'b'c']}+ \Gamma_{\lambda ca'}\Gamma_{ b'c'}^{\lambda}-\Gamma_{\lambda cb'}\Gamma_{ a'c'}^{\lambda}\label{tt29}
                            \end{eqnarray}

                                \begin{eqnarray}
                                && \bar{A}_{a'b'c'c}\rightarrow \bar{F}_{a'b'c'c}\rightarrow \bar{R}_{a'b'c'c} \nonumber \\
                                          &&  \bar{F}_{a_{1},a_{2},..a_{n}}=\partial_{[a_{5}}.. \partial_{a_{n}}\bar{R}_{a_{1},a_{2},a_{3},a_{4}
                                          ]}\label{tt30}
                                \end{eqnarray}
                                
  where ($g_{a'b'}$ and $\bar{g}_{a'b'}$) are graviton and gravitino respectively. Using equations of (\ref{t28},\ref{t29},\ref{t30},\ref{tt28},\ref{tt29},\ref{tt30}) in action of (\ref{P10} and \ref{PP10}) , we obtain the action of Gp-brane in terms of curvatures and metrics:

                                            \begin{eqnarray}
                                            &&S_{Gp} = -(T_{Gp}) \int dt \sum_{n=1}^{p}\beta_{n}\Big(
                                            \delta^{a_{1},a_{2}...a_{n}}_{b_{1}b_{2}....b_{n}}L^{b_{1}}_{a_{1}}...L^{b_{n}}_{a_{n}}\Big)^{1/2}\nonumber\\&&\nonumber\\&&\nonumber\\&&(L)^{a_{n}}_{b_{n},brane}=(L)^{a_{n}}_{b_{n},bosonic,brane}+(L)^{a_{n}}_{b_{n},fermionic,brane}\nonumber\\&&\nonumber\\&&\nonumber\\&&
                                            (L)^{a_{n}}_{b_{n},bosonic,brane}= \delta^{a_{n}}_{b_{n}}Tr\Big( 
                                               \frac{1}{2}i^{2(p-N)}\Sigma_{L=0}^{N} \Sigma_{H=0}^{N-L}\Sigma_{a_{1}..a_{L}=0}^{p}\Sigma_{j_{1}..j_{H}=p+1}^{M}(X^{j_{1}}..X^{j_{H-1}})^{2}\langle \partial_{a_{1}}..\partial_{a_{L}}X^{i},\partial_{a_{1}}..\partial_{a_{L}}X^{i}\rangle + \nonumber \\
                                                   && 
                                                          i^{2(p-N)} \Sigma_{L=0}^{N} \Sigma_{H=0}^{N-L}\Sigma_{a_{1}..a_{L}=0}^{p}\Sigma_{j_{1}..j_{H}=p+1}^{M}\frac{\lambda^{2}}{1.2...N}(X^{j_{1}}..X^{j_{H}})^{2}\langle \partial_{a_{5}}..\partial_{a_{L}}\tilde{R}^{a_{1},a_{2},a_{3},a_{4}},\partial_{a_{5}}..\partial_{a_{L}} \tilde{R}^{a_{1},a_{2},a_{3},a_{4}}\rangle+... \Big) \nonumber\\ &&
                                                            \nonumber \\
                                                                                                                    &&\nonumber \\
                                                                                                                    && (L)^{a_{n}}_{b_{n},fermionic,brane}= \delta^{a_{n}}_{b_{n}}Tr\Big( 
                                                                                                                                                         -\frac{1}{2}i^{2(p-N)+1}\Sigma_{L=0}^{N} \Sigma_{H=0}^{N-L}\Sigma_{a_{1}..a_{L}=0}^{p}\Sigma_{j_{1}..j_{H}=p+1}^{M}\times\nonumber \\
                                                                                                                                                                                                                   &&(X^{j_{1}}..X^{j_{H-1}})^{2}\gamma^{a_{L-1}}\langle \partial_{a_{1}}..\partial_{a_{L-1}}\psi^{i},\partial_{a_{1}}..\partial_{a_{L}}\psi^{i}\rangle -\nonumber \\
                                                                 &&  i^{2(p-N)+1}\Sigma_{L=0}^{N} \Sigma_{H=0}^{N-L}\Sigma_{a_{1}..a_{L}=0}^{p}\Sigma_{j_{1}..j_{H}=p+1}^{M}\frac{\lambda^{2}}{1.2...N}\gamma^{a_{L-1}}(X^{j_{1}}..X^{j_{H}})^{2}\langle \partial_{a_{5}}..\partial_{a_{L-1}}\bar{R}^{a_{1},a_{2},a_{3},a_{4}},\partial_{a_{5}}..\partial_{a_{L}} \bar{R}^{a_{1},a_{2},a_{3},a_{4}}\rangle +...\Big) \label{kp1}
                                            \end{eqnarray}
                         
                                            \begin{eqnarray}
                                              &&S_{Anti-Gp} = -(T_{Anti-Gp}) \int dt \sum_{n=1}^{p}\beta_{n}\Big(
                                              \delta^{a_{1},a_{2}...a_{n}}_{b_{1}b_{2}....b_{n}}L^{b_{1}}_{a_{1}}...L^{b_{n}}_{a_{n}}\Big)^{1/2}\nonumber\\&&\nonumber\\&&\nonumber\\&&(L)^{a_{n}}_{b_{n},anti-brane}=(L)^{a_{n}}_{b_{n},bosonic,anti-brane}+(L)^{a_{n}}_{b_{n},fermionic,anti-brane}\nonumber\\&&\nonumber\\&&\nonumber\\&&
                                              (L)^{a_{n}}_{b_{n},bosonic,anti-brane}= \delta^{a_{n}}_{b_{n}}Tr\Big( 
                                                 \frac{1}{2}i^{2(p-N+1)}\Sigma_{L=0}^{N} \Sigma_{H=0}^{N-L}\Sigma_{a_{1}..a_{L}=0}^{p}\Sigma_{j_{1}..j_{H}=p+1}^{M}(X^{j_{1}}..X^{j_{H-1}})^{2}\langle \partial_{a_{1}}..\partial_{a_{L}}X^{i},\partial_{a_{1}}..\partial_{a_{L}}X^{i}\rangle + \nonumber \\
                                                     && 
                                                            i^{2(p-N)} \Sigma_{L=0}^{N} \Sigma_{H=0}^{N-L}\Sigma_{a_{1}..a_{L}=0}^{p}\Sigma_{j_{1}..j_{H}=p+1}^{M}\frac{\lambda^{2}}{1.2...N}(X^{j_{1}}..X^{j_{H}})^{2}\langle \partial_{a_{5}}..\partial_{a_{L}}\hat{\tilde{R}}^{a_{1},a_{2},a_{3},a_{4}},\partial_{a_{5}}..\partial_{a_{L}} \hat{\tilde{R}}^{a_{1},a_{2},a_{3},a_{4}}\rangle+... \Big) \nonumber\\ &&
                                                              \nonumber \\
                                                                                                                      &&\nonumber \\
                                                                                                                      && (L)^{a_{n}}_{b_{n},fermionic,anti-brane}= \delta^{a_{n}}_{b_{n}}Tr\Big( 
                                                                                                                                                          - \frac{1}{2}i^{2(p-N+1)+1}\Sigma_{L=0}^{N} \Sigma_{H=0}^{N-L}\Sigma_{a_{1}..a_{L}=0}^{p}\Sigma_{j_{1}..j_{H}=p+1}^{M}\times\nonumber \\
                                                                                                                                                                                                                     &&(X^{j_{1}}..X^{j_{H-1}})^{2}\gamma^{a_{L-1}}\langle \partial_{a_{1}}..\partial_{a_{L-1}}\psi^{i},\partial_{a_{1}}..\partial_{a_{L}}\psi^{i}\rangle -\nonumber \\
                                                                   &&  i^{2(p-N)+1}\Sigma_{L=0}^{N} \Sigma_{H=0}^{N-L}\Sigma_{a_{1}..a_{L}=0}^{p}\Sigma_{j_{1}..j_{H}=p+1}^{M}\frac{\lambda^{2}}{1.2...N}\gamma^{a_{L-1}}(X^{j_{1}}..X^{j_{H}})^{2}\langle \partial_{a_{5}}..\partial_{a_{L-1}}\hat{\bar{R}}^{a_{1},a_{2},a_{3},a_{4}},\partial_{a_{5}}..\partial_{a_{L}} \hat{\bar{R}}^{a_{1},a_{2},a_{3},a_{4}}\rangle +...\Big) \label{kp2}
                                              \end{eqnarray}
                           
 In above actions, there exists  the high agreements between number of degrees of freedom for fermions and bosons like the usual actions in supergravity (\cite{h17,h18}). For example, for each scalar, there exists a fermionic superpatner and for various gravitonic terms with different orders of derivatives, variuos gravitinoic terms with the same orders of derivatives  appear.  In additional, the sign of bosonic gravity is different from  fermionic gravity  and they may cancel the effect of each other and produce the flat universe. On the other hand, with respect to the  difference in number of times on branes and anti-branes, different gravities may emerge which lead to difference in forces on branes and anti-branes.             
  
   \section{Emergence of the oscillating universe and wormholes in G-theory}\label{o2}
   
   Until now, we have obtained the shape of actions for  supergravity in G-theory. Now, we show that gravitons and gravitionoes produce two types of wormholes which prevent from complete contraction and expansion of branes and lead to the oscillation. To this end,  we use the method in (\cite{h4,h19}) and  calculate the momentum densities for bosonic part and fermionic part of Lagrangian in equations (\ref{kp1} and \ref{kp2} ) respect to derivatives of curvature. First, we start with derivatives of order of p-4, where p is dimension of brane and 4 denotes four indices of curvature. We get:

                         \begin{eqnarray}
                         && \Pi_{bosonic,brane,p-4}\approx \frac{ i^{2(p-N)} \Sigma_{L=0}^{N} \Sigma_{H=0}^{N-L}\Sigma_{a_{1}..a_{L}=0}^{p}\Sigma_{j_{1}..j_{H}=p+1}^{M}\frac{\lambda^{2}}{1.2...N}(X^{j_{1}}..X^{j_{H}})^{2}\partial_{a_{5}}..\partial_{a_{L}} \tilde{R}^{a_{1},a_{2},a_{3},a_{4}}}{\sqrt{(L)^{a_{n}}_{b_{n},bosonic,brane}}} \label{kp3}
                         \end{eqnarray}

                          \begin{eqnarray}
                          && \Pi_{fermionic,brane,p-4}\approx \frac{ -i^{2(p-N)+1} \Sigma_{L=0}^{N} \Sigma_{H=0}^{N-L}\Sigma_{a_{1}..a_{L}=0}^{p}\Sigma_{j_{1}..j_{H}=p+1}^{M}\frac{\lambda^{2}}{1.2...N}(X^{j_{1}}..X^{j_{H}})^{2}\partial_{a_{5}}..\partial_{a_{L}} \bar{R}^{a_{1},a_{2},a_{3},a_{4}}}{\sqrt{(L)^{a_{n}}_{b_{n},bosonic,brane}}} \label{kp4}
                          \end{eqnarray}

                             \begin{eqnarray}
                             && \Pi_{bosonic,anti-brane,p-4}\approx \frac{ i^{2(p-N+1)} \Sigma_{L=0}^{N} \Sigma_{H=0}^{N-L}\Sigma_{a_{1}..a_{L}=0}^{p}\Sigma_{j_{1}..j_{H}=p+1}^{M}\frac{\lambda^{2}}{1.2...N}(X^{j_{1}}..X^{j_{H}})^{2}\partial_{a_{5}}..\partial_{a_{L}} \bar{\tilde{R}}^{a_{1},a_{2},a_{3},a_{4}}}{\sqrt{(L)^{a_{n}}_{b_{n},bosonic,brane}}} \label{kp5}
                             \end{eqnarray}

                              \begin{eqnarray}
                              && \Pi_{fermionic,anti-brane,p-4}\approx \frac{ -i^{2(p-N+1)+1} \Sigma_{L=0}^{N} \Sigma_{H=0}^{N-L}\Sigma_{a_{1}..a_{L}=0}^{p}\Sigma_{j_{1}..j_{H}=p+1}^{M}\frac{\lambda^{2}}{1.2...N}(X^{j_{1}}..X^{j_{H}})^{2}\partial_{a_{5}}..\partial_{a_{L}} \hat{\bar{R}}^{a_{1},a_{2},a_{3},a_{4}}}{\sqrt{(L)^{a_{n}}_{b_{n},bosonic,brane}}} \label{kp6}
                              \end{eqnarray}

    We suppose that all coordinates are the same ($x^{2..p}=\sigma$) and build a p dimensional sphere. The Hamiltonian for the system of brane-anti-brane can be given by:

                          \begin{eqnarray}
                          &&H_{p-4}^{1}\approx 4\pi\int d\sigma \sigma^{p-1}\Sigma_{L=0}^{N} \Sigma_{H=0}^{N-L}\Sigma_{a_{1}..a_{L}=0}^{p}\Sigma_{j_{1}..j_{H}=p+1}^{M}\frac{\lambda^{2}}{1.2...N}(X^{j_{1}}..X^{j_{H}})^{2}\partial_{t}..\partial_{a_{L}} \tilde{R}^{a_{1},a_{2},a_{3},a_{4}} \Pi_{bosonic,brane,p-4}i^{2(p-N)} \nonumber\\&&
                                                    -4\pi\int d\sigma \sigma^{p-1} \Sigma_{L=0}^{N} \Sigma_{H=0}^{N-L}\Sigma_{a_{1}..a_{L}=0}^{p}\Sigma_{j_{1}..j_{H}=p+1}^{M}\frac{\lambda^{2}}{1.2...N}(X^{j_{1}}..X^{j_{H}})^{2}\partial_{t}..\partial_{a_{L}} \bar{R}^{a_{1},a_{2},a_{3},a_{4}}\Pi_{fermionic,brane,p-4}i^{2(p-N)+1} \nonumber\\&& +4\pi\int d\sigma \sigma^{p-1}\Sigma_{L=0}^{N} \Sigma_{H=0}^{N-L}\Sigma_{a_{1}..a_{L}=0}^{p}\Sigma_{j_{1}..j_{H}=p+1}^{M}\frac{\lambda^{2}}{1.2...N}(X^{j_{1}}..X^{j_{H}})^{2}\partial_{t}..\partial_{a_{L}} \hat{\tilde{R}}^{a_{1},a_{2},a_{3},a_{4}} \Pi_{bosonic,anti-brane,p-4}i^{2(p-N+1)}\nonumber\\&&-4\pi\int d\sigma \sigma^{p-1} \Sigma_{L=0}^{N} \Sigma_{H=0}^{N-L}\Sigma_{a_{1}..a_{L}=0}^{p}\Sigma_{j_{1}..j_{H}=p+1}^{M}\frac{\lambda^{2}}{1.2...N}(X^{j_{1}}..X^{j_{H}})^{2}\partial_{t}..\partial_{a_{L}} \hat{\bar{R}}^{a_{1},a_{2},a_{3},a_{4}}\Pi_{fermionic,anti-brane,p-4}i^{2(p-N+1)+1} \nonumber\\&&-L_{bosonic,brane,p-4}^{1}-L_{fermionic,brane,p-4}^{1}\nonumber\\&&-L_{bosonic,anti-brane,p-4}^{1}-L_{fermionic,anti-brane,p-4}^{1}=\nonumber\\&&4\pi\int d\sigma \sigma^{p-1}\Sigma_{L=0}^{N} \Sigma_{H=0}^{N-L}\Sigma_{a_{1}..a_{L}=0}^{p}\Sigma_{j_{1}..j_{H}=p+1}^{M}\frac{\lambda^{2}}{1.2...N}(X^{j_{1}}..X^{j_{H}})^{2}\partial_{a_{5}}..\partial_{a_{L}} \tilde{R}^{a_{1},a_{2},a_{3},a_{4}} \Pi_{bosonic,brane,p-4}i^{2(p-N)} \nonumber\\&&
                          -4\pi\int d\sigma \sigma^{p-1} \Sigma_{L=0}^{N} \Sigma_{H=0}^{N-L}\Sigma_{a_{1}..a_{L}=0}^{p}\Sigma_{j_{1}..j_{H}=p+1}^{M}\frac{\lambda^{2}}{1.2...N}(X^{j_{1}}..X^{j_{H}})^{2}\partial_{a_{5}}..\partial_{a_{L}} \bar{R}^{a_{1},a_{2},a_{3},a_{4}}\Pi_{fermionic,brane,p-4}i^{2(p-N)+1} \nonumber\\&& +4\pi\int d\sigma \sigma^{p-1}\Sigma_{L=0}^{N} \Sigma_{H=0}^{N-L}\Sigma_{a_{1}..a_{L}=0}^{p}\Sigma_{j_{1}..j_{H}=p+1}^{M}\frac{\lambda^{2}}{1.2...N}(X^{j_{1}}..X^{j_{H}})^{2}\partial_{a_{5}}..\partial_{a_{L}} \hat{\tilde{R}}^{a_{1},a_{2},a_{3},a_{4}} \Pi_{bosonic,anti-brane,p-4}i^{2(p-N+1)} \nonumber\\&&+4\pi\int d\sigma  \Sigma_{L=0}^{N} \Sigma_{H=0}^{N-L}\Sigma_{a_{1}..a_{L}=0}^{p}\Sigma_{j_{1}..j_{H}=p+1}^{M}\frac{\lambda^{2}}{1.2...N}(X^{j_{1}}..X^{j_{H}})^{2}\partial_{a_{5}}..\partial_{a_{L}} \hat{\bar{R}}^{a_{1},a_{2},a_{3},a_{4}}\Pi_{fermionic,anti-brane,p-4}i^{2(p-N+1)+1} \nonumber\\&&-4\pi\int d\sigma \sigma^{p-1}\Sigma_{L=0}^{N} \Sigma_{H=0}^{N-L}\Sigma_{a_{1}..a_{L}=0}^{p}\Sigma_{j_{1}..j_{H}=p+1}^{M}\frac{\lambda^{2}}{1.2...N}(X^{j_{1}}..X^{j_{H}})^{2}\partial_{a_{6}}..\partial_{a_{L}} \tilde{R}^{a_{1},a_{2},a_{3},a_{4}}\times\nonumber\\&&\partial_{a_{5}}(\sigma^{p-1} \Pi_{bosonic,brane,p-4}i^{2(p-N)}) \nonumber\\&&
                                                    -4\pi\int d\sigma  \Sigma_{L=0}^{N} \Sigma_{H=0}^{N-L}\Sigma_{a_{1}..a_{L}=0}^{p}\Sigma_{j_{1}..j_{H}=p+1}^{M}\frac{\lambda^{2}}{1.2...N}(X^{j_{1}}..X^{j_{H}})^{2}\partial_{a_{6}}..\partial_{a_{L}} \bar{R}^{a_{1},a_{2},a_{3},a_{4}}\times\nonumber\\&&\partial_{a_{5}}(\sigma^{p-1}\Pi_{fermionic,brane,p-4}i^{2(p-N)+1} )\nonumber\\&& -4\pi\int d\sigma \Sigma_{L=0}^{N} \Sigma_{H=0}^{N-L}\Sigma_{a_{1}..a_{L}=0}^{p}\Sigma_{j_{1}..j_{H}=p+1}^{M}\frac{\lambda^{2}}{1.2...N}(X^{j_{1}}..X^{j_{H}})^{2}\partial_{a_{6}}..\partial_{a_{L}} \hat{\tilde{R}}^{a_{1},a_{2},a_{3},a_{4}}\times\nonumber\\&& \partial_{a_{5}}(\sigma^{p-1}\Pi_{bosonic,anti-brane,p-4}i^{2(p-N+1)}) \nonumber\\&&+4\pi\int d\sigma  \Sigma_{L=0}^{N} \Sigma_{H=0}^{N-L}\Sigma_{a_{1}..a_{L}=0}^{p}\Sigma_{j_{1}..j_{H}=p+1}^{M}\frac{\lambda^{2}}{1.2...N}(X^{j_{1}}..X^{j_{H}})^{2}\partial_{a_{6}}..\partial_{a_{L}} \hat{\bar{R}}^{a_{1},a_{2},a_{3},a_{4}}\times\nonumber\\&&\partial_{a_{5}}(\sigma^{p-1}\Pi_{fermionic,anti-brane,p-4}i^{2(p-N+1)+1})  \nonumber\\&&-L_{bosonic,brane,p-4}^{1}-L_{fermionic,brane,p-4}^{1}\nonumber\\&&-L_{bosonic,anti-brane,p-4}^{1}-L_{fermionic,anti-brane,p-4}^{1}
                          \label{kp7}
                          \end{eqnarray}

     Here, we have applied in the second step integrated by parts. We can use the constraint ($\partial_{a_{5}}(\sigma^{p-1}\Pi_{bosonic/fermionic,brane/anti-brane,p-4}i^{2(p-N)})/i^{2(p-N+1)+1}) =0$) and derive the momentum densities:

                                 \begin{eqnarray}
                                 && \Pi_{bosonic,brane,p-4}=\frac{i^{2(p-N)}k_{bosonic,brane,p-4}}{\sigma^{p-1}}\nonumber\\&&\Pi_{fermionic,brane,p-4}=-\frac{i^{2(p-N)+1}k_{fermionic,brane,p-4}}{\sigma^{p-1}}\nonumber\\&& \Pi_{bosonic,anti-brane,p-4}=\frac{i^{2(p-N+1)}k_{bosonic,anti-brane,p-4}}{\sigma^{p-1}}\nonumber\\&&\Pi_{fermionic,anti-brane,p-4}=-\frac{i^{2(p-N+1)+1}k_{fermionic,anti-brane,p-4}}{\sigma^{p-1}}\label{kp8}
                                 \end{eqnarray}  
                                                                             
 Substituting momentum densities of equation (\ref{kp8})  in equation (\ref{kp7}), we can obtain the Hamiltonian as:

                                  \begin{eqnarray}
                                  && 
                                   H_{p-4}^{1}\approx 4\pi\int d\sigma \sigma^{p-1}\Big([ \frac{1}{2}i^{2(p-N)}\Sigma_{L=0}^{N} \Sigma_{H=0}^{N-L}\Sigma_{a_{1}..a_{L}=0}^{p}\Sigma_{j_{1}..j_{H}=p+1}^{M}(X^{j_{1}}..X^{j_{H-1}})^{2}\langle \partial_{a_{1}}..\partial_{a_{L}}X^{i},\partial_{a_{1}}..\partial_{a_{L}}X^{i}\rangle + \nonumber \\
                                                                             && 
                                                                                    i^{2(p-N)} \Sigma_{L=0}^{N} \Sigma_{H=0}^{N-L}\Sigma_{a_{1}..a_{L}=0}^{p}\Sigma_{j_{1}..j_{H}=p+1}^{M}\frac{\lambda^{2}}{1.2...N}(X^{j_{1}}..X^{j_{H}})^{2}\langle \partial_{a_{6}}..\partial_{a_{L}}\tilde{R}^{a_{1},a_{2},a_{3},a_{4}},\partial_{a_{6}}..\partial_{a_{L}} \tilde{R}^{a_{1},a_{2},a_{3},a_{4}}\rangle+... ]^{1/2}\times\nonumber \\
                                                                                                                                                                 &&\sqrt{1+(\frac{i^{2(p-N)}k_{bosonic,brane,p-4}}{\sigma^{p-1}})^{2}}+\nonumber \\&&
                                                                                                           [-\frac{1}{2}i^{2(p-N)+1}\Sigma_{L=0}^{N} \Sigma_{H=0}^{N-L}\Sigma_{a_{1}..a_{L}=0}^{p}\Sigma_{j_{1}..j_{H}=p+1}^{M}(X^{j_{1}}..X^{j_{H-1}})^{2}\gamma^{a_{L-1}}\langle \partial_{a_{1}}..\partial_{a_{L-1}}\psi^{i},\partial_{a_{1}}..\partial_{a_{L}}\psi^{i}\rangle -\nonumber \\
                                                                                                                                                                   &&  i^{2(p-N)+1}\Sigma_{L=0}^{N} \Sigma_{H=0}^{N-L}\Sigma_{a_{1}..a_{L}=0}^{p}\Sigma_{j_{1}..j_{H}=p+1}^{M}\frac{\lambda^{2}}{1.2...N}\gamma^{a_{L-1}}(X^{j_{1}}..X^{j_{H}})^{2}\langle \partial_{a_{6}}..\partial_{a_{L-1}}\bar{R}^{a_{1},a_{2},a_{3},a_{4}},\partial_{a_{6}}..\partial_{a_{L}} \bar{R}^{a_{1},a_{2},a_{3},a_{4}}\rangle +...]^{1/2} \times  \nonumber \\&& \sqrt{1+(-\frac{i^{2(p-N)+1}k_{fermionic,brane,p-4}}{\sigma^{p-1}})^{2}}+ \nonumber \\&& [ \frac{1}{2}i^{2(p-N+1)}\Sigma_{L=0}^{N} \Sigma_{H=0}^{N-L}\Sigma_{a_{1}..a_{L}=0}^{p}\Sigma_{j_{1}..j_{H}=p+1}^{M}(X^{j_{1}}..X^{j_{H-1}})^{2}\langle \partial_{a_{1}}..\partial_{a_{L}}X^{i},\partial_{a_{1}}..\partial_{a_{L}}X^{i}\rangle + \nonumber \\
                                                                                                                                                                                                               && 
                                                                                                                                                                                                                      i^{2(p-N)} \Sigma_{L=0}^{N} \Sigma_{H=0}^{N-L}\Sigma_{a_{1}..a_{L}=0}^{p}\Sigma_{j_{1}..j_{H}=p+1}^{M}\frac{\lambda^{2}}{1.2...N}(X^{j_{1}}..X^{j_{H}})^{2}\langle \partial_{a_{6}}..\partial_{a_{L}}\hat{\tilde{R}}^{a_{1},a_{2},a_{3},a_{4}},\partial_{a_{6}}..\partial_{a_{L}} \hat{\tilde{R}}^{a_{1},a_{2},a_{3},a_{4}}\rangle+... ]^{1/2}\times\nonumber \\&&\sqrt{1+(\frac{i^{2(p-N+1)}k_{bosonic,anti-brane,p-4}}{\sigma^{p-1}})^{2}}                                                    +\nonumber \\&& [  - \frac{1}{2}i^{2(p-N+1)+1}\Sigma_{L=0}^{N} \Sigma_{H=0}^{N-L}\Sigma_{a_{1}..a_{L}=0}^{p}\Sigma_{j_{1}..j_{H}=p+1}^{M}(X^{j_{1}}..X^{j_{H-1}})^{2}\gamma^{a_{L-1}}\langle \partial_{a_{1}}..\partial_{a_{L-1}}\psi^{i},\partial_{a_{1}}..\partial_{a_{L}}\psi^{i}\rangle -\nonumber \\
                                                                                                                                                                                                                                                                                &&  i^{2(p-N)+1}\Sigma_{L=0}^{N} \Sigma_{H=0}^{N-L}\Sigma_{a_{1}..a_{L}=0}^{p}\Sigma_{j_{1}..j_{H}=p+1}^{M}\frac{\lambda^{2}}{1.2...N}\gamma^{a_{L-1}}(X^{j_{1}}..X^{j_{H}})^{2}\langle \partial_{a_{6}}..\partial_{a_{L-1}}\hat{\bar{R}}^{a_{1},a_{2},a_{3},a_{4}},\partial_{a_{6}}..\partial_{a_{L}} \hat{\bar{R}}^{a_{1},a_{2},a_{3},a_{4}}\rangle +..]^{1/2}\times\nonumber \\                                                                                                                                                                                                                                                                                                                                                                                                                                                                                                                                                                                                                                                                                                               &&\sqrt{1+(-\frac{i^{2(p-N+1)+1}k_{fermionic,anti-brane,p-4}}{\sigma^{p-1}})^{2}}\Big)
                                   \label{kp9}
                                  \end{eqnarray}  
                                  
  We apply the previous method again and calculate momentum densitis for curvature of order p-5: 
  
                           \begin{eqnarray}
                           && \Pi_{bosonic,brane,p-5}\approx \frac{ i^{2(p-N)} \Sigma_{L=0}^{N} \Sigma_{H=0}^{N-L}\Sigma_{a_{1}..a_{L}=0}^{p}\Sigma_{j_{1}..j_{H}=p+1}^{M}\frac{\lambda^{2}}{1.2...N}(X^{j_{1}}..X^{j_{H}})^{2}\partial_{a_{6}}..\partial_{a_{L}} \tilde{R}^{a_{1},a_{2},a_{3},a_{4}}}{H_{p-4}^{1}} \label{kp10}
                           \end{eqnarray}

                            \begin{eqnarray}
                            && \Pi_{fermionic,brane,p-5}\approx \frac{ -i^{2(p-N)+1} \Sigma_{L=0}^{N} \Sigma_{H=0}^{N-L}\Sigma_{a_{1}..a_{L}=0}^{p}\Sigma_{j_{1}..j_{H}=p+1}^{M}\frac{\lambda^{2}}{1.2...N}(X^{j_{1}}..X^{j_{H}})^{2}\partial_{a_{6}}..\partial_{a_{L}} \bar{R}^{a_{1},a_{2},a_{3},a_{4}}}{H_{p-4}^{1}} \label{kp11}
                            \end{eqnarray}

                               \begin{eqnarray}
                               && \Pi_{bosonic,anti-brane,p-5}\approx \frac{ i^{2(p-N+1)} \Sigma_{L=0}^{N} \Sigma_{H=0}^{N-L}\Sigma_{a_{1}..a_{L}=0}^{p}\Sigma_{j_{1}..j_{H}=p+1}^{M}\frac{\lambda^{2}}{1.2...N}(X^{j_{1}}..X^{j_{H}})^{2}\partial_{a_{6}}..\partial_{a_{L}} \bar{\tilde{R}}^{a_{1},a_{2},a_{3},a_{4}}}{H_{p-4}^{1}} \label{kp12}
                               \end{eqnarray}

                                \begin{eqnarray}
                                && \Pi_{fermionic,anti-brane,p-5}\approx \frac{ -i^{2(p-N+1)+1} \Sigma_{L=0}^{N} \Sigma_{H=0}^{N-L}\Sigma_{a_{1}..a_{L}=0}^{p}\Sigma_{j_{1}..j_{H}=p+1}^{M}\frac{\lambda^{2}}{1.2...N}(X^{j_{1}}..X^{j_{H}})^{2}\partial_{a_{6}}..\partial_{a_{L}} \hat{\bar{R}}^{a_{1},a_{2},a_{3},a_{4}}}{H_{p-4}^{1}} \label{kp13}
                                \end{eqnarray}

 We substitute derivatives of order p-5 instead of these momentums and  get the Hamiltonian as follows:

                          \begin{eqnarray}
                          &&H_{p-5}^{1}\approx 4\pi\int d\sigma \sigma^{p-1}\Sigma_{L=0}^{N} \Sigma_{H=0}^{N-L}\Sigma_{a_{1}..a_{L}=0}^{p}\Sigma_{j_{1}..j_{H}=p+1}^{M}\frac{\lambda^{2}}{1.2...N}(X^{j_{1}}..X^{j_{H}})^{2}\partial_{a_{5}}..\partial_{a_{L}} \tilde{R}^{a_{1},a_{2},a_{3},a_{4}} \Pi_{bosonic,brane,p-5}i^{2(p-N)}\times\nonumber\\&&\sqrt{1+(\frac{i^{2(p-N)}k_{bosonic,brane,p-4}}{\sigma^{p-1}})^{2}} \nonumber\\&&
                          -4\pi\int d\sigma \sigma^{p-1} \Sigma_{L=0}^{N} \Sigma_{H=0}^{N-L}\Sigma_{a_{1}..a_{L}=0}^{p}\Sigma_{j_{1}..j_{H}=p+1}^{M}\frac{\lambda^{2}}{1.2...N}(X^{j_{1}}..X^{j_{H}})^{2}\partial_{a_{5}}..\partial_{a_{L}} \bar{R}^{a_{1},a_{2},a_{3},a_{4}}\Pi_{fermionic,brane,p-5}i^{2(p-N)+1}\times\nonumber\\&&\sqrt{1+(\frac{i^{2(p-N)+1}k_{fermionic,brane,p-4}}{\sigma^{p-1}})^{2}} \nonumber\\&& +4\pi\int d\sigma \sigma^{p-1}\Sigma_{L=0}^{N} \Sigma_{H=0}^{N-L}\Sigma_{a_{1}..a_{L}=0}^{p}\Sigma_{j_{1}..j_{H}=p+1}^{M}\frac{\lambda^{2}}{1.2...N}(X^{j_{1}}..X^{j_{H}})^{2}\partial_{a_{5}}..\partial_{a_{L}} \hat{\tilde{R}}^{a_{1},a_{2},a_{3},a_{4}} \Pi_{bosonic,anti-brane,p-5}i^{2(p-N+1)}\times\nonumber\\&&\sqrt{1+(\frac{i^{2(p-N+1)}k_{bosonic,anti-brane,p-4}}{\sigma^{p-1}})^{2}} \nonumber\\&&+4\pi\int d\sigma  \Sigma_{L=0}^{N} \Sigma_{H=0}^{N-L}\Sigma_{a_{1}..a_{L}=0}^{p}\Sigma_{j_{1}..j_{H}=p+1}^{M}\frac{\lambda^{2}}{1.2...N}(X^{j_{1}}..X^{j_{H}})^{2}\partial_{a_{5}}..\partial_{a_{L}} \hat{\bar{R}}^{a_{1},a_{2},a_{3},a_{4}}\Pi_{fermionic,anti-brane,p-5}i^{2(p-N+1)+1}\times\nonumber\\&&\sqrt{1+(\frac{i^{2(p-N+1)+1}k_{fermionic,anti-brane,p-4}}{\sigma^{p-1}})^{2}} \nonumber\\&&-4\pi\int d\sigma \sigma^{p-1}\Sigma_{L=0}^{N} \Sigma_{H=0}^{N-L}\Sigma_{a_{1}..a_{L}=0}^{p}\Sigma_{j_{1}..j_{H}=p+1}^{M}\frac{\lambda^{2}}{1.2...N}(X^{j_{1}}..X^{j_{H}})^{2}\partial_{a_{6}}..\partial_{a_{L}} \tilde{R}^{a_{1},a_{2},a_{3},a_{4}}\times\nonumber\\&&\partial_{a_{6}}(\sqrt{1+(\frac{i^{2(p-N)}k_{bosonic,brane,p-4}}{\sigma^{p-1}})^{2}}\sigma^{p-1} \Pi_{bosonic,brane,p-5}i^{2(p-N)}) \nonumber\\&&
                                                    -4\pi\int d\sigma  \Sigma_{L=0}^{N} \Sigma_{H=0}^{N-L}\Sigma_{a_{1}..a_{L}=0}^{p}\Sigma_{j_{1}..j_{H}=p+1}^{M}\frac{\lambda^{2}}{1.2...N}(X^{j_{1}}..X^{j_{H}})^{2}\partial_{a_{6}}..\partial_{a_{L}} \bar{R}^{a_{1},a_{2},a_{3},a_{4}}\times\nonumber\\&&\partial_{a_{6}}(\sqrt{1+(\frac{i^{2(p-N)+1}k_{fermionic,brane,p-4}}{\sigma^{p-1}})^{2}}\sigma^{p-1}\Pi_{fermionic,brane,p-5}i^{2(p-N)+1} )\nonumber\\&& -4\pi\int d\sigma \Sigma_{L=0}^{N} \Sigma_{H=0}^{N-L}\Sigma_{a_{1}..a_{L}=0}^{p}\Sigma_{j_{1}..j_{H}=p+1}^{M}\frac{\lambda^{2}}{1.2...N}(X^{j_{1}}..X^{j_{H}})^{2}\partial_{a_{6}}..\partial_{a_{L}} \hat{\tilde{R}}^{a_{1},a_{2},a_{3},a_{4}}\times\nonumber\\&& \partial_{a_{6}}(\sqrt{1+(\frac{i^{2(p-N+1)}k_{bosonic,anti-brane,p-4}}{\sigma^{p-1}})^{2}}\sigma^{p-1}\Pi_{bosonic,anti-brane,p-5}i^{2(p-N+1)}) \nonumber\\&&+4\pi\int d\sigma  \Sigma_{L=0}^{N} \Sigma_{H=0}^{N-L}\Sigma_{a_{1}..a_{L}=0}^{p}\Sigma_{j_{1}..j_{H}=p+1}^{M}\frac{\lambda^{2}}{1.2...N}(X^{j_{1}}..X^{j_{H}})^{2}\partial_{a_{6}}..\partial_{a_{L}} \hat{\bar{R}}^{a_{1},a_{2},a_{3},a_{4}}\times\nonumber\\&&\partial_{a_{6}}(\sqrt{1+(\frac{i^{2(p-N+1)+1}k_{fermionic,anti-brane,p-4}}{\sigma^{p-1}})^{2}}\sigma^{p-1}\Pi_{fermionic,anti-brane,p-5}i^{2(p-N+1)+1})  \nonumber\\&&-L_{bosonic,brane,p-4}^{1}-L_{fermionic,brane,p-4}^{1}\nonumber\\&&-L_{bosonic,anti-brane,p-4}^{1}-L_{fermionic,anti-brane,p-4}^{1}
                          \label{kp14}
                          \end{eqnarray}  
 Similar to previous stage, we impose the constraints ($\partial_{a_{6}}(\sqrt{1+(\frac{i^{.....}k}{\sigma^{p-1}})^{2}}\sigma^{p-1}\Pi i^{..})=0 $) and derive the below momentums:

                                  \begin{eqnarray}
                                  && \Pi_{bosonic,brane,p-5}=\frac{i^{2(p-N)}k_{bosonic,brane,p-5}}{\sigma^{p-1}\sqrt{1+(\frac{i^{2(p-N)}k_{bosonic,brane,p-4}}{\sigma^{p-1}})^{2}}}\nonumber\\&&\Pi_{fermionic,brane,p-4}=-\frac{i^{2(p-N)+1}k_{fermionic,brane,p-4}}{\sigma^{p-1}\sqrt{1+(\frac{i^{2(p-N)+1}k_{fermionic,brane,p-4}}{\sigma^{p-1}})^{2}}}\nonumber\\&& \Pi_{bosonic,anti-brane,p-5}=\frac{i^{2(p-N+1)}k_{bosonic,anti-brane,p-5}}{\sigma^{p-1}\sqrt{1+(\frac{i^{2(p-N+1)}k_{bosonic,anti-brane,p-4}}{\sigma^{p-1}})^{2}}}\nonumber\\&&\Pi_{fermionic,anti-brane,p-5}=-\frac{i^{2(p-N+1)+1}k_{fermionic,anti-brane,p-5}}{\sigma^{p-1}\sqrt{1+(\frac{i^{2(p-N+1)+1}k_{fermionic,anti-brane,p-4}}{\sigma^{p-1}})^{2}}}\label{kp15}
                                  \end{eqnarray}  
                                        
  Substituting these momentums in Hamiltonian (\ref{kp14}), we obtain the below Hamiltonian:

                                    \begin{eqnarray}
                                    && 
                                     H_{p-4}^{1}\approx 4\pi\int d\sigma \sigma^{p-1}\Big([ \frac{1}{2}i^{2(p-N)}\Sigma_{L=0}^{N} \Sigma_{H=0}^{N-L}\Sigma_{a_{1}..a_{L}=0}^{p}\Sigma_{j_{1}..j_{H}=p+1}^{M}(X^{j_{1}}..X^{j_{H-1}})^{2}\langle \partial_{a_{1}}..\partial_{a_{L}}X^{i},\partial_{a_{1}}..\partial_{a_{L}}X^{i}\rangle + \nonumber \\
                                                                               && 
                                                                                      i^{2(p-N)} \Sigma_{L=0}^{N} \Sigma_{H=0}^{N-L}\Sigma_{a_{1}..a_{L}=0}^{p}\Sigma_{j_{1}..j_{H}=p+1}^{M}\frac{\lambda^{2}}{1.2...N}(X^{j_{1}}..X^{j_{H}})^{2}\langle \partial_{a_{7}}..\partial_{a_{L}}\tilde{R}^{a_{1},a_{2},a_{3},a_{4}},\partial_{a_{7}}..\partial_{a_{L}} \tilde{R}^{a_{1},a_{2},a_{3},a_{4}}\rangle+... ]^{1/2}\times\nonumber \\
                                                                                                                                                                   &&\sqrt{1+(\frac{i^{2(p-N)}k_{bosonic,brane,p-5}}{\sigma^{p-1}\sqrt{1+(\frac{i^{2(p-N)}k_{bosonic,brane,p-4}}{\sigma^{p-1}})^{2}}})^{2}}+\nonumber \\&&
                                                                                                             [-\frac{1}{2}i^{2(p-N)+1}\Sigma_{L=0}^{N} \Sigma_{H=0}^{N-L}\Sigma_{a_{1}..a_{L}=0}^{p}\Sigma_{j_{1}..j_{H}=p+1}^{M}(X^{j_{1}}..X^{j_{H-1}})^{2}\gamma^{a_{L-1}}\langle \partial_{a_{1}}..\partial_{a_{L-1}}\psi^{i},\partial_{a_{1}}..\partial_{a_{L}}\psi^{i}\rangle -\nonumber \\
                                                                                                                                                                     &&  i^{2(p-N)+1}\Sigma_{L=0}^{N} \Sigma_{H=0}^{N-L}\Sigma_{a_{1}..a_{L}=0}^{p}\Sigma_{j_{1}..j_{H}=p+1}^{M}\frac{\lambda^{2}}{1.2...N}\gamma^{a_{L-1}}(X^{j_{1}}..X^{j_{H}})^{2}\langle \partial_{a_{7}}..\partial_{a_{L-1}}\bar{R}^{a_{1},a_{2},a_{3},a_{4}},\partial_{a_{7}}..\partial_{a_{L}} \bar{R}^{a_{1},a_{2},a_{3},a_{4}}\rangle +...]^{1/2} \times  \nonumber \\&& \sqrt{1+(-\frac{i^{2(p-N)+1}k_{fermionic,brane,p-5}}{\sigma^{p-1}\sqrt{1+(-\frac{i^{2(p-N)+1}k_{fermionic,brane,p-4}}{\sigma^{p-1}})^{2}}})^{2}}+ \nonumber \\&& [ \frac{1}{2}i^{2(p-N+1)}\Sigma_{L=0}^{N} \Sigma_{H=0}^{N-L}\Sigma_{a_{1}..a_{L}=0}^{p}\Sigma_{j_{1}..j_{H}=p+1}^{M}(X^{j_{1}}..X^{j_{H-1}})^{2}\langle \partial_{a_{1}}..\partial_{a_{L}}X^{i},\partial_{a_{1}}..\partial_{a_{L}}X^{i}\rangle + \nonumber \\
                                                                                                                                                                                                                 && 
                                                                                                                                                                                                                        i^{2(p-N)} \Sigma_{L=0}^{N} \Sigma_{H=0}^{N-L}\Sigma_{a_{1}..a_{L}=0}^{p}\Sigma_{j_{1}..j_{H}=p+1}^{M}\frac{\lambda^{2}}{1.2...N}(X^{j_{1}}..X^{j_{H}})^{2}\langle \partial_{a_{6}}..\partial_{a_{L}}\hat{\tilde{R}}^{a_{1},a_{2},a_{3},a_{4}},\partial_{a_{6}}..\partial_{a_{L}} \hat{\tilde{R}}^{a_{1},a_{2},a_{3},a_{4}}\rangle+... ]^{1/2}\times\nonumber \\&&\sqrt{1+(\frac{i^{2(p-N+1)}k_{bosonic,anti-brane,p-5}}{\sigma^{p-1}\sqrt{1+(\frac{i^{2(p-N+1)}k_{bosonic,anti-brane,p-4}}{\sigma^{p-1}})^{2}}})^{2}}                                                    +\nonumber \\&& [  - \frac{1}{2}i^{2(p-N+1)+1}\Sigma_{L=0}^{N} \Sigma_{H=0}^{N-L}\Sigma_{a_{1}..a_{L}=0}^{p}\Sigma_{j_{1}..j_{H}=p+1}^{M}(X^{j_{1}}..X^{j_{H-1}})^{2}\gamma^{a_{L-1}}\langle \partial_{a_{1}}..\partial_{a_{L-1}}\psi^{i},\partial_{a_{1}}..\partial_{a_{L}}\psi^{i}\rangle -\nonumber \\
                                                                                                                                                                                                                                                                                  &&  i^{2(p-N)+1}\Sigma_{L=0}^{N} \Sigma_{H=0}^{N-L}\Sigma_{a_{1}..a_{L}=0}^{p}\Sigma_{j_{1}..j_{H}=p+1}^{M}\frac{\lambda^{2}}{1.2...N}\gamma^{a_{L-1}}(X^{j_{1}}..X^{j_{H}})^{2}\langle \partial_{a_{6}}..\partial_{a_{L-1}}\hat{\bar{R}}^{a_{1},a_{2},a_{3},a_{4}},\partial_{a_{6}}..\partial_{a_{L}} \hat{\bar{R}}^{a_{1},a_{2},a_{3},a_{4}}\rangle +..]^{1/2}\times\nonumber \\                                                                                                                                                                                                                                                                                                                                                                                                                                                                                                                                                                                                                                                                                                               &&\sqrt{1+(-\frac{i^{2(p-N+1)+1}k_{fermionic,anti-brane,p-5}}{\sigma^{p-1}\sqrt{1+(-\frac{i^{2(p-N+1)+1}k_{fermionic,anti-brane,p-4}}{\sigma^{p-1}})^{2}}})^{2}}\Big)
                                     \label{kp16}
                                    \end{eqnarray}  
    
After doing some algebra, we can replace all derivatives respect to curvatures by function (F) and obtain the Hamiltonian as follows:

                                    \begin{eqnarray}
                                    && 
                                     H_{tot}^{1}\approx 4\pi\int d\sigma \sigma^{p-1} \Big([ \frac{1}{2}i^{2(p-N)}\Sigma_{L=0}^{N} \Sigma_{H=0}^{N-L}\Sigma_{a_{1}..a_{L}=0}^{p}\Sigma_{j_{1}..j_{H}=p+1}^{M}(X^{j_{1}}..X^{j_{H-1}})^{2}\langle \partial_{a_{1}}..\partial_{a_{L}}X^{i},\partial_{a_{1}}..\partial_{a_{L}}X^{i}\rangle + ... ]^{1/2}\times\nonumber \\
                                                                                                                                                                   &&F_{bosonic,brane,tot}+\nonumber \\&&
                                                                                                             [-\frac{1}{2}i^{2(p-N)+1}\Sigma_{L=0}^{N} \Sigma_{H=0}^{N-L}\Sigma_{a_{1}..a_{L}=0}^{p}\Sigma_{j_{1}..j_{H}=p+1}^{M}(X^{j_{1}}..X^{j_{H-1}})^{2}\gamma^{a_{L-1}}\langle \partial_{a_{1}}..\partial_{a_{L-1}}\psi^{i},\partial_{a_{1}}..\partial_{a_{L}}\psi^{i}\rangle +...]^{1/2} \times  \nonumber \\&&F_{fermionic,brane,tot}+ \nonumber \\&& [ \frac{1}{2}i^{2(p-N+1)}\Sigma_{L=0}^{N} \Sigma_{H=0}^{N-L}\Sigma_{a_{1}..a_{L}=0}^{p}\Sigma_{j_{1}..j_{H}=p+1}^{M}(X^{j_{1}}..X^{j_{H-1}})^{2}\langle \partial_{a_{1}}..\partial_{a_{L}}X^{i},\partial_{a_{1}}..\partial_{a_{L}}X^{i}\rangle +... ]^{1/2}\times\nonumber \\&&F_{bosonic,anti-brane,tot}                                                   +\nonumber \\&& [  - \frac{1}{2}i^{2(p-N+1)+1}\Sigma_{L=0}^{N} \Sigma_{H=0}^{N-L}\Sigma_{a_{1}..a_{L}=0}^{p}\Sigma_{j_{1}..j_{H}=p+1}^{M}(X^{j_{1}}..X^{j_{H-1}})^{2}\gamma^{a_{L-1}}\langle \partial_{a_{1}}..\partial_{a_{L-1}}\psi^{i},\partial_{a_{1}}..\partial_{a_{L}}\psi^{i}\rangle +..]^{1/2}\times\nonumber \\                                                                                                                                                                                                                                                                                                                                                                                                                                                                                                                                                                                                                                                                                                               &&F_{fermionic,anti-brane,tot}\Big)
                                     \label{kp17}
                                    \end{eqnarray}  
 where functions of $F$  are defined as follows:

                                   \begin{eqnarray}
                                   && F_{bosonic,brane,tot}= \sqrt{1+(\frac{i^{2(p-N)}k_{bosonic,brane,1}}{\sigma^{p-1}\sqrt{1+(\frac{i^{2(p-N)}k_{bosonic,brane,2}}{\sigma^{p-1}\sqrt{1+(\frac{i^{2(p-N)}k_{bosonic,brane,3}}{\sigma^{p-1}...\sqrt{1+(\frac{i^{2(p-N)}k_{bosonic,brane,p-4}}{\sigma^{p-1}})^{2}}})^{2}}})^{2}}})^{2}}\nonumber \\                                  &&F_{fermionic,brane,tot}= \sqrt{1+(\frac{i^{2(p-N)+1}k_{fermionic,brane,1}}{\sigma^{p-1}\sqrt{1+(\frac{i^{2(p-N)+1}k_{fermionic,brane,2}}{\sigma^{p-1}\sqrt{1+(\frac{i^{2(p-N)+1}k_{fermionic,brane,3}}{\sigma^{p-1}...\sqrt{1+(\frac{i^{2(p-N)+1}k_{fermionic,brane,p-4}}{\sigma^{p-1}})^{2}}})^{2}}})^{2}}})^{2}}\nonumber \\                                                                                                                                                                                                                                                                                                                                                                                                                                                                                                                                                                                                                                                                                                               &&F_{bosonic,anti-brane,tot}= \sqrt{1+(\frac{i^{2(p-N+1)}k_{bosonic,anti-brane,1}}{\sigma^{p-1}\sqrt{1+(\frac{i^{2(p-N+1)}k_{bosonic,anti-brane,2}}{\sigma^{p-1}\sqrt{1+(\frac{i^{2(p-N+1)}k_{bosonic,anti-brane,3}}{\sigma^{p-1}...\sqrt{1+(\frac{i^{2(p-N+1)}k_{bosonic,anti-brane,p-4}}{\sigma^{p-1}})^{2}}})^{2}}})^{2}}})^{2}}\nonumber \\                                                                                                                                                                                                                                                                                                                                                                                                                                                                                                                                                                                                                                                                                                               &&F_{fermionic,anti-brane,tot}= \sqrt{1+(\frac{i^{2(p-N+1)+1}k_{fermionic,anti-brane,1}}{\sigma^{p-1}\sqrt{1+(\frac{i^{2(p-N+1)+1}k_{fermionic,anti-brane,2}}{\sigma^{p-1}\sqrt{1+(\frac{i^{2(p-N+1)+1}k_{fermionic,anti-brane,3}}{\sigma^{p-1}...\sqrt{1+(\frac{i^{2(p-N+1)+1}k_{fermionic,anti-brane,p-4}}{\sigma^{p-1}})^{2}}})^{2}}})^{2}}})^{2}}\label{kp18}
                                   \end{eqnarray} 
                                    
   These Hamiltonians are very similar to the Hamiltonians of BIon in \cite{h2,h3,h4,h19}. Obviously,  curvatures of bosonic gravitons and fermionic gravitinoes create two types of wormhole which their signatures and couplings arenot the same. The sign of Hamiltonians of wormholes which are produced by bosonic gravitons and fermionic gravitinoes on anti-branes is different. This means that the potential energy of one brane which is  negative, produces attractive force and the potential energy of another brane which is positive produces the repulsive force. Thus, physics of branes is different from the one of anti-branes.  Now, for simplicity, we assume  $x^{0}=it,x^{1,2,3}=\sigma$, $X^{0}=t$,$X^{1}=z_{+}+iz_{-}$, $X^{i}=0,i\neq 0,1$, $\psi^{0}=t$,  $\psi^{1}=y_{+}+iy_{-}$,  $\psi^{i}=0,i\neq 0,1$, $i^{2(p-N)}=1$ and $\gamma^{a_{L-1}}=i^{2(N-L-1)-1}$ where indices $\pm$ refer to the fields on brane and anti-branes respectively. Also, $\sigma$ is the separation distance between two ends of one brane, z and y denote the lengths of bosonic and fermionic wormholes between branes and their nth derivatives are shown by $z^{n(')},y^{n(')}$. Using these assumption in equation (\ref{kp17}), we re-obtain Hamiltonian of the system of branes-anti-branes as:

                                        \begin{eqnarray}
                                        && 
                                         H_{tot}^{1}\approx 4\pi\int d\sigma \sigma^{p-1} \Big([1+ \Sigma_{n=1}^{N-1}(z_{+}^{(N-n-1)}z_{+}^{n(')})^{2} ]^{1/2}F_{bosonic,brane,tot}+\nonumber \\&&
                                                                                                                 [-1+\Sigma_{n=1}^{N-1}(-iz_{+})^{2(N-n-1)}y_{+}^{n}y_{+}^{n(')}]^{1/2} F_{fermionic,brane,tot}+ \nonumber \\&& [-1+ i^{2N-2}\Sigma_{n=1}^{N-1}(z_{-}^{(N-n-1)}z_{-}^{n(')})^{2} ]^{1/2}F_{bosonic,anti-brane,tot}                                                   +\nonumber \\&& [1+i^{2N-2}\Sigma_{n=1}^{N-1}(-iz_{-})^{2(N-n-1)}y_{-}^{n}y_{-}^{n(')}]^{1/2}F_{fermionic,anti-brane,tot}\Big)
                                         \label{kp19}
                                        \end{eqnarray} 
   
   Now, we can calculate the equation of motion for $z$ and y:

                                        \begin{eqnarray}
                                        && 
                                        \Big(\Sigma_{n=1}^{N-1}(-1)^{n}(z_{+}^{(N-n-1)})^{2}z_{+}^{(n)(')}z_{+}^{(n-1)(')}\frac{\sigma^{p-1}F_{bosonic,brane,tot}}{[1+ \Sigma_{n=1}^{N-1}(z_{+}^{(N-n-1)}z_{+}^{n(')})^{2} ]^{1/2}}\Big)'=\nonumber\\&&\Big(\Sigma_{n=1}^{N-1}(z_{+}^{n(')})^{2}z_{+}^{2(N-n-1)-1}\frac{\sigma^{p-1}F_{bosonic,brane,tot}}{[1+ \Sigma_{n=1}^{N-1}(z_{+}^{(N-n-1)}z_{+}^{n(')})^{2} ]^{1/2}}\Big)
                                         \label{kp20}
                                        \end{eqnarray}

                                          \begin{eqnarray}
                                          && 
                                          \Big(\Sigma_{n=1}^{N-1}(-1)^{n}(-iz_{+})^{2(N-n-1)}y_{+}^{n}y_{+}^{(n-1)(')}\frac{\sigma^{p-1}F_{fermionic,brane,tot}}{ [-1+\Sigma_{n=1}^{N-1}(-iz_{+})^{2(N-n-1)}y_{+}^{n}y_{+}^{n(')}]^{1/2}}\Big)'=\nonumber\\&&\Big(\Sigma_{n=1}^{N-1}(-iz_{+})^{2(N-n-1)}y_{+}^{n-1}y_{+}^{n(')}\frac{\sigma^{p-1}F_{fermionic,brane,tot}}{ [-1+\Sigma_{n=1}^{N-1}(-iz_{+})^{2(N-n-1)}y_{+}^{n}y_{+}^{n(')}]^{1/2}}\Big)
                                           \label{kp21}
                                          \end{eqnarray}

                                           \begin{eqnarray}
                                           && 
                                           \Big(\Sigma_{n=1}^{N-1}i^{2N-2}(-1)^{n}(z_{-}^{(N-n-1)})^{2}z_{-}^{(n)(')}z_{-}^{(n-1)(')}\frac{\sigma^{p-1}F_{bosonic,anti-brane,tot}}{[-1+ i^{2N-2}\Sigma_{n=1}^{N-1}(z_{-}^{(N-n-1)}z_{-}^{n(')})^{2} ]^{1/2}}\Big)'=\nonumber\\&&\Big(\Sigma_{n=1}^{N-1}i^{2N-2}(z_{-}^{n(')})^{2}z_{+}^{2(N-n-1)-1}\frac{\sigma^{p-1}F_{bosonic,anti-brane,tot}}{[-1+ i^{2N-2}\Sigma_{n=1}^{N-1}(z_{-}^{(N-n-1)}z_{-}^{n(')})^{2} ]^{1/2} }\Big)
                                            \label{kp22}
                                           \end{eqnarray}

                                             \begin{eqnarray}
                                             && 
                                             \Big(\Sigma_{n=1}^{N-1}i^{2N-2}(-1)^{n}(-iz_{-})^{2(N-n-1)}y_{-}^{n}y_{-}^{(n-1)(')}\frac{\sigma^{p-1}F_{fermionic,anti-brane,tot}}{ [1+i^{2N-2}\Sigma_{n=1}^{N-1}(-iz_{-})^{2(N-n-1)}y_{-}^{n}y_{-}^{n(')}]^{1/2}}\Big)'=\nonumber\\&&\Big(\Sigma_{n=1}^{N-1}i^{2N-2}(-iz_{-})^{2(N-n-1)}y_{-}^{n-1}y_{-}^{n(')}\frac{\sigma^{p-1}F_{fermionic,anti-brane,tot}}{ [1+i^{2N-2}\Sigma_{n=1}^{N-1}(-iz_{-})^{2(N-n-1)}y_{-}^{n}y_{-}^{n(')}]^{1/2}}\Big)
                                              \label{kp23}
                                             \end{eqnarray}

                                             Solving these equations, we obtain:

                                          \begin{eqnarray}
                                          && z_{+}=z_{+,0}\Sigma_{n=0}^{N-1}e^{-\int d^{n}\sigma F_{bosonic,brane,tot}(\sigma)\frac{1}{F_{bosonic,brane,tot}^{-n}(\sigma_{0,bosonic,brane})-F_{bosonic,brane,tot}^{-n}(\sigma)}}  \Big(1+\nonumber\\&&
                                         \int d^{n}\sigma F_{bosonic,brane,tot}^{-1}(\sigma)(F_{bosonic,brane,tot}^{-n}(\sigma_{0,bosonic,brane})-F_{bosonic,brane,tot}^{-n}(\sigma))\sin(n\sigma)  \Big)\label{kp24}
                                          \end{eqnarray}

                                          \begin{eqnarray}
                                          && y_{+}=y_{+,0}\Sigma_{n=0}^{N-1}e^{-\int d^{n}\sigma F_{fermionic,brane,tot}^{-1}(\sigma)\frac{1}{F_{fermionic,brane,tot}^{n}(\sigma)-F_{fermionic,brane,tot}^{n}(\sigma_{0,fermionic,brane})}}  \Big(1+\nonumber\\&&
                                         \int d^{n}\sigma F_{fermionic,brane,tot}(\sigma)(F_{fermionic,brane,tot}^{n}(\sigma)-F_{fermionic,brane,tot}^{n}(\sigma_{0,fermionic,brane}))\cos(n\sigma)  \Big)\label{kp25}
                                          \end{eqnarray} 
                                                          
     \begin{eqnarray}
                                           && z_{-}=z_{-,0}\Sigma_{n=0}^{N-1}e^{-\int d^{n}\sigma F_{bosonic,anti-brane,tot}(\sigma)\frac{1}{F_{bosonic,anti-brane,tot}^{-n}(\sigma)-F_{bosonic,anti-brane,tot}^{-n}(\sigma_{0,bosonic,anti-brane})}}  \Big(1+\nonumber\\&&
                                          \int d^{n}\sigma F_{bosonic,anti-brane,tot}^{-1}(\sigma)(F_{bosonic,anti-brane,tot}^{-n}(\sigma)-F_{bosonic,anti-brane,tot}^{-n}(\sigma_{0,bosonic,anti-brane}))\sin(n\sigma)  \Big)\label{kp26}
                                           \end{eqnarray}

                                           \begin{eqnarray}
                                           && y_{-}=y_{-,0}\Sigma_{n=0}^{N-1}e^{-\int d^{n}\sigma F_{fermionic,anti-brane,tot}^{-1}(\sigma)\frac{1}{F_{fermionic,anti-brane,tot}^{n}(\sigma_{0,fermionic,anti-brane})-F_{bosonic,anti-brane,tot}^{n}(\sigma)}}  \Big(1+\nonumber\\&&
                                          \int d^{n}\sigma F_{fermionic,anti-brane,tot}(\sigma)(F_{fermionic,anti-brane,tot}^{n}(\sigma_{0,fermionic,anti-brane})-F_{fermionic,anti-brane,tot}^{n}(\sigma))\cos(n\sigma)  \Big)\label{kp27}
                                           \end{eqnarray} 
                                                                                                                          where  $\sigma_{0}$ is throat of wormhole. To build universe in BIon, we assume that ($\sigma$) is the separation distance between ends of one brane. Thus, these solutions show that for contacting point ( $\sigma=0$), the length of gravitonic wormholes is zero. By increasing the separation distance between ends of brane ($\sigma$), this length increases, turns over a maximum and tends to zero at throat $\sigma_{0}$. Then, one new bosonic wormhole is produced, it's length grows  with increasing $\sigma$ and tends to infinity at $\sigma=\infty$. On the other hand, the length of fermionic wormholes is $\infty$ at contracting point ( $\sigma=0$ ) and tends to zero at throat $\sigma_{0}$. Then one new fermionic wormhole is born, it's length increases with increasing $\sigma$, turns over a maximum and tends to zero when the separation distance between two ends of one brane is $\infty$. Thus, fermionic and bosonic wormholes  prevent from closing and getting away of ends of branes from each other and lead to the oscillation.
                                                                                                                          
                                                                                                                          Applying equations (\ref{kp24},\ref{kp25},\ref{kp26},\ref{kp27}) in equation (\ref{kp19}), we calculate the potential for this system:

                                         \begin{eqnarray}
                                         && H_{tot}\approx V_{tot}=V_{bosonic,brane,tot}+V_{fermionic,brane,tot}+V_{bosonic,anti-brane,tot}+V_{fermionic,anti-brane,tot}\nonumber\\&&\nonumber\\&&\nonumber\\&&
                                          V_{bosonic,brane,tot}\approx 4\pi\int d\sigma \Big(\Big( \Sigma_{n'=1}^{N-1}[z_{+,0}\Sigma_{n=0}^{N-1}e^{-\int d^{n}\sigma F_{bosonic,brane,tot}(\sigma)\frac{1}{F_{bosonic,brane,tot}^{-n}(\sigma_{0,bosonic,brane})-F_{bosonic,brane,tot}^{-n}(\sigma)}} \nonumber\\&& \Big(1+
                                                                                   \int d^{n}\sigma F_{bosonic,brane,tot}^{-1}(\sigma)(F_{bosonic,brane,tot}^{-n}(\sigma_{0,bosonic,brane})-F_{bosonic,brane,tot}^{-n}(\sigma))\sin(n\sigma)  \Big)]^{2N-2n'-2}\times
   \nonumber\\&&[F_{bosonic,brane,tot}^{-1}(\sigma)(F_{bosonic,brane,tot}^{-n}(\sigma_{0,bosonic,brane})-F_{bosonic,brane,tot}^{-n}(\sigma))\sin(n\sigma)\times\nonumber\\&& e^{-\int d^{n}\sigma F_{bosonic,brane,tot}(\sigma)\frac{1}{F_{bosonic,brane,tot}^{-n}(\sigma_{0,bosonic,brane})-F_{bosonic,brane,tot}^{-n}(\sigma)}}+\nonumber\\&&e^{-\int d^{n}\sigma F_{bosonic,brane,tot}(\sigma)\frac{1}{F_{bosonic,brane,tot}^{-n}(\sigma_{0,bosonic,brane})-F_{bosonic,brane,tot}^{-n}(\sigma)}}\times\nonumber\\&&(F_{bosonic,brane,tot}(\sigma)\frac{1}{F_{bosonic,brane,tot}^{-n}(\sigma_{0,bosonic,brane})-F_{bosonic,brane,tot}^{-n}(\sigma)})]^{2}F_{bosonic,brane,tot}(\sigma) \Big)\Big)\nonumber\\&&\nonumber\\&&\nonumber\\&&
                                             V_{fermionic,brane,tot}\approx 4\pi\int d\sigma \Big(\Big( \Sigma_{n'=1}^{N-1}[z_{+,0}\Sigma_{n=0}^{N-1}e^{-\int d^{n}\sigma F_{bosonic,brane,tot}(\sigma)\frac{1}{F_{bosonic,brane,tot}^{-n}(\sigma_{0,bosonic,brane})-F_{bosonic,brane,tot}^{-n}(\sigma)}} \nonumber\\&& \Big(1+
                                                                                      \int d^{n}\sigma F_{bosonic,brane,tot}^{-1}(\sigma)(F_{bosonic,brane,tot}^{-n}(\sigma_{0,bosonic,brane})-F_{bosonic,brane,tot}^{-n}(\sigma))\sin(n\sigma)\Big) ]^{2N-2n'-2}\times\nonumber\\&&[y_{+,0}\Sigma_{n=0}^{N-1}e^{-\int d^{n}\sigma F_{fermionic,brane,tot}^{-1}(\sigma)\frac{1}{F_{fermionic,brane,tot}^{n}(\sigma)-F_{fermionic,brane,tot}^{n}(\sigma_{0,fermionic,brane})}}  \Big(1+\nonumber\\&&
                                                                                                                               \int d^{n}\sigma F_{fermionic,brane,tot}(\sigma)(F_{fermionic,brane,tot}^{n}(\sigma)-F_{fermionic,brane,tot}^{n}(\sigma_{0,fermionic,brane}))\cos(n\sigma)  \Big)]^{n'}\times\nonumber\\&&[y_{+,0}\Sigma_{n=0}^{N-1}(e^{-\int d^{n}\sigma F_{fermionic,brane,tot}^{-1}(\sigma)\frac{1}{F_{fermionic,brane,tot}^{n}(\sigma)-F_{fermionic,brane,tot}^{n}(\sigma_{0,fermionic,brane})}}\times  \nonumber\\&&
                                                                                                                                                                         F_{fermionic,brane,tot}(\sigma)(F_{fermionic,brane,tot}^{n}(\sigma)-F_{fermionic,brane,tot}^{n}(\sigma_{0,fermionic,brane}))\cos(n\sigma))+ \nonumber\\&& \Big(e^{-\int d^{n}\sigma F_{fermionic,brane,tot}^{-1}(\sigma)\frac{1}{F_{fermionic,brane,tot}^{n}(\sigma)-F_{fermionic,brane,tot}^{n}(\sigma_{0,fermionic,brane})}}\times  \nonumber\\&&F_{fermionic,brane,tot}^{-1}(\sigma)\frac{1}{F_{fermionic,brane,tot}^{n}(\sigma)-F_{fermionic,brane,tot}^{n}(\sigma_{0,fermionic,brane})}) \Big)]F_{fermionic,brane,tot}\Big)\Big)\nonumber\\&&\nonumber\\&&\nonumber\\&&
                                                                                                                                                                                                                   V_{bosonic,anti-brane,tot}\nonumber\\&& \approx 4\pi\int d\sigma \Big(\Big( \Sigma_{n'=1}^{N-1}[z_{-,0}\Sigma_{n=0}^{N-1}e^{-\int d^{n}\sigma F_{bosonic,anti-brane,tot}(\sigma)\frac{1}{F_{bosonic,anti-brane,tot}^{-n}(\sigma)-F_{bosonic,brane,tot}^{-n}(\sigma_{0,anti-bosonic,brane})}} \nonumber\\&& \Big(1+
                                                                                                                                                                                                                                                            \int d^{n}\sigma F_{bosonic,anti-brane,tot}^{-1}(\sigma)(F_{bosonic,anti-brane,tot}^{-n}(\sigma)-F_{bosonic,brane,tot}^{-n}(\sigma_{0,anti-bosonic,brane}))\sin(n\sigma)  \Big)]^{2N-2n'-2}\times
                                                                                                                                                                            \nonumber\\&&[F_{bosonic,anti-brane,tot}^{-1}(\sigma)(F_{bosonic,anti-brane,tot}^{-n}(\sigma)-F_{bosonic,anti-brane,tot}^{-n}(\sigma_{0,anti-bosonic,brane}))\sin(n\sigma)\times\nonumber\\&& e^{-\int d^{n}\sigma F_{bosonic,anti-brane,tot}(\sigma)\frac{1}{F_{bosonic,anti-brane,tot}^{-n}(\sigma)-F_{bosonic,anti-brane,tot}^{-n}(\sigma_{0,anti-bosonic,brane})}}+\nonumber\\&&e^{-\int d^{n}\sigma F_{bosonic,anti-brane,tot}(\sigma)\frac{1}{F_{bosonic,anti-brane,tot}^{-n}(\sigma)-F_{bosonic,anti-brane,tot}^{-n}(\sigma_{0,anti-bosonic,brane})}}\times\nonumber\\&&(F_{bosonic,anti-brane,tot}(\sigma)\frac{1}{F_{bosonic,anti-brane,tot}^{-n}(\sigma)-F_{bosonic,anti-brane,tot}^{-n}(\sigma_{0,anti-bosonic,brane})})]^{2}\times\nonumber\\&&F_{bosonic,anti-brane,tot}(\sigma) \Big)\Big)\nonumber\\&&\nonumber\\&&\nonumber\\&&                                                                        V_{fermionic, anti-brane,tot}\approx \nonumber\\&& 4\pi\int d\sigma \Big(\Big( \Sigma_{n'=1}^{N-1}[z_{-,0}\Sigma_{n=0}^{N-1}e^{-\int d^{n}\sigma F_{bosonic,anti-brane,tot}(\sigma)\frac{1}{F_{bosonic,anti-brane,tot}^{-n}(\sigma)-F_{bosonic,anti-brane,tot}^{-n}(\sigma_{0,bosonic,anti-brane})}} \nonumber\\&& \Big(1+
                                                                                                                                                                                                                                                                  \int d^{n}\sigma F_{bosonic,anti-brane,tot}^{-1}(\sigma)(F_{bosonic,anti-brane,tot}^{-n}(\sigma)-F_{bosonic,anti-brane,tot}^{-n}(\sigma_{0,bosonic,anti-brane}))\times\nonumber\\&&\sin(n\sigma)\Big) ]^{2N-2n'-2}[y_{-,0}\Sigma_{n=0}^{N-1}\times\nonumber\\&&e^{-\int d^{n}\sigma F_{fermionic,anti-brane,tot}^{-1}(\sigma)\frac{1}{F_{fermionic,anti-brane,tot}^{n}(\sigma_{0,fermionic,anti-brane})-F_{fermionic,anti-brane,tot}^{n}(\sigma)}}\nonumber\\&&  \Big(1+
                                                                                                                                                                                                                                                                                                           \int d^{n}\sigma F_{fermionic,anti-brane,tot}(\sigma)(F_{fermionic,anti-brane,tot}^{n}(\sigma_{0,fermionic,anti-brane})-\nonumber\\&& F_{fermionic,anti-anti-brane,tot}^{n}(\sigma))\cos(n\sigma)  \Big)]^{n'}\times\nonumber\\&&[y_{-,0}\Sigma_{n=0}^{N-1}(e^{-\int d^{n}\sigma F_{fermionic,anti-brane,tot}^{-1}(\sigma)\frac{1}{F_{fermionic,anti-brane,tot}^{n}(\sigma_{0,fermionic,anti-brane})-F_{fermionic,anti-brane,tot}^{n}(\sigma)}}\times  \nonumber\\&&
                                                                                                                                                                                                                                                                                                                                                     F_{fermionic,anti-brane,tot}(\sigma)(F_{fermionic,anti-brane,tot}^{n}(\sigma_{0,fermionic,anti-brane})-F_{fermionic,anti-brane,tot}^{n}(\sigma))\cos(n\sigma))+ \nonumber\\&& \Big(e^{-\int d^{n}\sigma F_{fermionic,anti-brane,tot}^{-1}(\sigma)\frac{1}{F_{fermionic,anti-brane,tot}^{n}(\sigma_{0,fermionic,anti-brane})-F_{fermionic,anti-brane,tot}^{n}(\sigma)}}\times  \nonumber\\&&F_{fermionic,anti-brane,tot}^{-1}(\sigma)\frac{1}{F_{fermionic,anti-brane,tot}^{n}(\sigma_{0,fermionic,anti-brane})-F_{fermionic,anti-brane,tot}^{n}(\sigma)}) \Big)]\times\nonumber\\&&F_{fermionic,anti-brane,tot}\Big)\Big)      
                                          \label{kp28}
                                         \end{eqnarray} 
                                                                                                                          For simplicity, we suppose that $\sigma_{0,bosonic,brane}=\sigma_{0,bosonic,anti-brane}$ and  $\sigma_{0 fermionic,brane}=\sigma_{0,fermionic,anti-brane}$ and calculate the  potentials and their relative forces between particles and anti-particles approximately:

                                                                                                                                                                     \begin{eqnarray}
                                             &&   V_{tot}=V_{bosonic,brane+anti-brane,tot}+V_{fermionic,brane+anti-brane,tot}\nonumber\\&&\nonumber\\&&\nonumber\\&& V_{bosonic,brane+anti-brane,tot}\approx 1-\Sigma_{m=1}^{P-1}\Sigma_{n=1}^{N-1} [k_{bosonic,brane,}-k_{bosonic,anti-brane}]^{m}\sigma^{m}\Big(\sigma_{0,bosonic,brane}^{nm}-\sigma^{nm}\Big)\nonumber\\&&\nonumber\\&& \text{ For $\sigma \ll \sigma_{0,bosonic,brane}$} \quad F_{bosonic}\approx\Sigma_{m=1}^{P-1}\Sigma_{n=1}^{N-1} [k_{bosonic,brane,}-k_{bosonic,anti-brane}]^{m}m\sigma^{m-1}\sigma_{0,bosonic,brane}^{nm}\nonumber\\&& \text{ For $\sigma \gg \sigma_{0,bosonic,brane}$} \quad F_{bosonic}\approx-\Sigma_{m=1}^{P-1}\Sigma_{n=1}^{N-1} [k_{bosonic,brane,}-k_{bosonic,anti-brane}]^{m}n(m+1)\sigma^{m}\sigma^{nm-1}\nonumber\\&&\nonumber\\&&\nonumber\\&&
                                             V_{fermionic,brane+anti-brane,tot}\approx \Sigma_{m=1}^{P-1}\Sigma_{n=1}^{N-1} [k_{fermionic,brane,}-k_{fermionic,anti-brane}]^{m}\frac{1}{\sigma^{m}}\Big(\sigma^{-nm}-\sigma^{-nm}_{0,fermionic,brane}\Big) \nonumber\\&&\nonumber\\&& \text{ For $\sigma \ll \sigma_{0,fermionic,brane}$} \quad F_{fermionic}\approx \Sigma_{m=1}^{P-1}\Sigma_{n=1}^{N-1} [k_{fermionic,brane,}-k_{fermionic,anti-brane}]^{m}\frac{m+nm}{\sigma^{m+nm+1}}\nonumber\\&& \text{ For $\sigma \gg \sigma_{0,fermionic,brane}$} \quad F_{fermionic}\approx-\Sigma_{m=1}^{P-1}\Sigma_{n=1}^{N-1} [k_{fermionic,brane,}-k_{fermionic,anti-brane}]^{m}\frac{m+1}{\sigma^{m+1}\sigma^{nm}_{0,fermionic,brane}}                               \label{kp29}
                                                                                                                                                                    \end{eqnarray}

 These results show that when two ends of one brane are completely close to each other at contracting point( $\sigma=0$), the potential of gravitonic wormholes is zero. By growing the separation distance between ends of branes and it's expansion, bosonic wormhole creates a repulsive potential and anti-gravity force which first increases, turns over a maximum and then reduces to zero at $\sigma_{0,bosonic,brane}$. After this distance, bosonic potential becomes attractive, gravity is born and prevents from getting away of ends of one brane. On the other hand, the gravitino creates a wormhole which leads to production of repulsive potential and  anti-gravity for small separation distance between ends of one brane. This potential is $\infty $ at contracting point  ( $\sigma=0$) and causes that two ends of one brane get away from each other fastly. By growing the separation distance between these ends, repulsive gravity decreases and tends to zero at $\sigma_{0,fermionic,brane}$. Then, the sign of potential changes and anti-gravity converts to gravity. This gravity increases, turns over a maximum and tends to zero when separation distance between two ends of one brane is $\infty $. 
 
 To study the evolution of universe in G-theory, we
 introduce two four dimensional universes that interact with each
 other via two bosonic and fermionic wormholes and build a BIonic system. In this model, our
  universe is placed on one Gp-brane and connected by another
 universe on the anti-Gp-brane by two fermionic and bosonic wormholes.  With respect to
 the FRW metric,
 
 \begin{eqnarray}
 && ds^{2}_{FRW}  = -dt^{2} + a(t)^{2}(dx^{2} +
 dy^{2} + dz^{2}), \label{a4}
 \end{eqnarray}
 
 we can calculate the energy density and momentum for one flat
 universe\cite{h2,h17,h18,h20}:
 
 \begin{eqnarray}
 && \rho_{uni} = 3 H^{2},
  \nonumber \\ && p_{uni} = H^{2} + 2\frac{\ddot{a}}{a} \label{a5}
 \end{eqnarray}
 
 Where $H=\frac{\dot{a}}{a}$ is the Hubble parameter and a is the scale factor.  The metric of two bosonic and fermionic wormholes
 are given by \cite{h20}:
 \begin{eqnarray}
 && ds_{wormhole,boson}^{2} =
 -dt^{2}+\frac{Q_{boson}^{2}(t)}{1-b(r)}dr^{2}+Q_{boson}^{2}(t)r^{2}d\phi^{2}\nonumber\\&& ds_{wormhole,fermion}^{2} =
  -dt^{2}+\frac{Q_{fermion}^{2}(t)}{1-b(r)}dr^{2}+Q_{fermion}^{2}(t)r^{2}d\phi^{2}\label{a6}
 \end{eqnarray}
 
where for $b(r)\ll Q^{2}(t)$ , we obtain the energy density and momentum as:                 

 \begin{eqnarray}
 && \rho_{wormhole,boson} =  \frac{M(M-1)\dot{Q}^{2}_{boson}}{2Q_{boson}^{2}} \nonumber\\&& p_{wormhole,fermion} = -  \frac{(M-2)(M-1)\dot{Q}^{2}_{boson}}{2Q_{boson}^{2}}- \frac{(M-1)\ddot{Q}_{boson}}{Q_{boson}} \nonumber\\&& \rho_{wormhole,boson} =  \frac{M(M-1)\dot{Q}^{2}_{fermion}}{2Q_{fermion}^{2}} \nonumber\\&& p_{wormhole,fermion} = -  \frac{(M-2)(M-1)\dot{Q}^{2}_{fermion}}{2Q_{fermion}^{2}}- \frac{(M-1)\ddot{Q}_{fermion}}{Q_{fermion}} \label{aa5}
 \end{eqnarray}
                                                                                                           where M is number of dimensions. According to conversation law, the energy density of two universes in additional to two bosonic and fermionic wormholes should be equal to energy density of BIon. We can write:
                                                                                                                                                                                                                                                                                                                                   \begin{eqnarray}
  &&  \rho_{BIon}=\rho_{uni1}+\rho_{uni2}+ \rho_{wormhole,boson}+ \rho_{wormhole,fermion}   \nonumber\\&&
     p_{BIon}=p_{uni1}+p_{uni2}+ p_{wormhole,boson}+ p_{wormhole,fermion}                                                                                                        \label{aaa5}
                                                                                                             \end{eqnarray}
                                                                                                           By assuming, $\sigma=t$ and $\sigma_{0}=t_{0}$, where $t_{0}$ is boundary time between contraction and expansion phase, the energy density and momentum are given by:
                                                                                                                                                                                                                     \begin{eqnarray}                                                                                             &&  \rho_{BIon}\approx V_{tot}=V_{bosonic,brane+anti-brane,tot}+V_{fermionic,brane+anti-brane,tot}\approx  \nonumber\\&&  1-\Sigma_{m=1}^{P-1}\Sigma_{n=1}^{N-1} [k_{bosonic,brane,}-k_{bosonic,anti-brane}]^{m}t^{m}\Big(t_{0,bosonic,brane}^{nm}-t^{nm}\Big)+ \nonumber\\&& \Sigma_{m=1}^{P-1}\Sigma_{n=1}^{N-1} [k_{fermionic,brane,}-k_{fermionic,anti-brane}]^{m}\frac{1}{t^{m}}\Big(t^{-nm}-t^{-nm}_{0,fermionic,brane}\Big)\nonumber\\&&\nonumber\\&&\nonumber\\&&                                                                                                              p_{BIon}= \int dt F= -\int dt \frac{\partial V_{tot}}{\partial \sigma} = -\int dt \frac{\partial V_{tot}}{\partial t}\approx-V_{tot}\approx -\rho_{BIon}                                                                                                    \label{ta5}
  \end{eqnarray} 
  
  Using equations (\ref{a5},\ref{aa5},\ref{aaa5} and \ref{ta5}) and also assuming ($\ddot{Q}\ll Q$) and M=4, we can obtain the scale factor of universe and parameters of wormholes in terms of time:
  
\begin{eqnarray}                                                                                             &&  a(t)\approx a_{0} e^{-1+\frac{1}{2}\Sigma_{m=1}^{P-1}\Sigma_{n=1}^{N-1} [k_{bosonic,brane,}-k_{bosonic,anti-brane}]^{m}t^{m}\Big(t_{0,bosonic,brane}^{nm}-t^{nm}\Big)}\times \nonumber\\&& e^{-\frac{1}{2}\Sigma_{m=1}^{P-1}\Sigma_{n=1}^{N-1} [k_{fermionic,brane,}-k_{fermionic,anti-brane}]^{m}\frac{1}{t^{m}}\Big(t^{-nm}-t^{-nm}_{0,fermionic,brane}\Big)}\nonumber\\&& \nonumber\\&&\nonumber\\&&\nonumber\\&&                                                                                                              Q_{boson}\approx e^{-\frac{1}{2}\Sigma_{m=1}^{P-1}\Sigma_{n=1}^{N-1} [k_{bosonic,brane,}-k_{bosonic,anti-brane}]^{m}t^{m}\Big(t_{0,bosonic,brane}^{nm}-t^{nm}\Big)}\times \nonumber\\&& [\frac{1}{2}\Sigma_{m=1}^{P-1}\Sigma_{n=1}^{N-1} [k_{bosonic,brane,}-k_{bosonic,anti-brane}]^{m}t^{m}\Big(t_{0,bosonic,brane}^{nm}-t^{nm}\Big)]^{2}\nonumber\\&&\nonumber\\&&\nonumber\\&& Q_{fermion}\approx e^{\frac{1}{2}\Sigma_{m=1}^{P-1}\Sigma_{n=1}^{N-1} [k_{fermionic,brane,}-k_{fermionic,anti-brane}]^{m}\frac{1}{t^{m}}\Big(t^{-nm}-t^{-nm}_{0,fermionic,brane}\Big)} \times \nonumber\\&& [\frac{1}{2}\Sigma_{m=1}^{P-1}\Sigma_{n=1}^{N-1} [k_{fermionic,brane,}-k_{fermionic,anti-brane}]^{m}\frac{1}{t^{m}}\Big(t^{-nm}-t^{-nm}_{0,fermionic,brane}\Big)]^{2} \label{ta5}
  \end{eqnarray}
  
  where $a_{0}$ is the scale factor of universe between contaction and expansion phases and is the maximal length of universe. These equations show that scale factor of universe is zero at ($t=0$) which is a maximal of contraction branch. Then, it grows by passing time and tends to $a=a_{0}$  at $t=t_{s}$ (end of expansion phase) and  after this point, shrinks to zero at $t=\infty$ (end of contraction epoch). These evolutions are controlled by bosonic and fermionic wormholes. At maximal contacting point (t=0), the parameter of fermionic wormhole is infinite and this wormhole produces a repulsive gravity which leads to opening universe, ending contraction era and starting expansion branch. During expansion phase, the parameter of bosonic wormhole grows and tend to infinity. This wormhole creates an attractive gravity which leads to contracting universe and end of expansion phase. Thus, in G-theory, all universes, wormholes and fields are emerged from nothing and Big Bang puzzle can be resolved completely.
    
  \section{ Horava-Witten mechanism in G-theory  }\label{o3}
  In this section, we will show that CGG terms which is applied for removing the anomaly in Horawa-Witten mechanism  \cite{b1,b2} can be riginated from higher dimensions. Also, the physics of D-dimensional manifold + Lie-N-algebra=the physics of D+N-dimensional manifold. For example, the physics of 11-dimensional manifold in Horava-Witten mechanism \cite{b1,b2} + Lie-three-algebra= the physics of 14-dimensional manifold \cite{A1}. For this, we will argue  that by adding one three dimensional manifold to 11-dimensional manifold in supergravity, GG terms in the action make a transition  to CGG terms. In fact, 11-dimensional anomalies may be removed in 14-dimensions completely. 
  First, we begin with  the bosonic terms in the eleven dimensional supergravity model \cite{b1,b2}:
       
   \begin{eqnarray}
   && S_{Bosonic-SUGRA} = \frac{1}{\bar{\kappa}^{2}}\int d^{11}x\sqrt{g}\Big(-\frac{1}{2}R-\frac{1}{48}G_{IJKL}G^{IJKL}\Big) + S_{CKK} \nonumber\\&& S_{CGG}=-\frac{\sqrt{2}}{3456\bar{\kappa}^{2}}\int_{M^{11}}d^{11}x \varepsilon^{I_{1}I_{2}...I_{11}}C_{I_{1}I_{2}I_{3}}G_{I_{4}...I_{7}}G_{I_{8}...I_{11}}  \label{s1}
   \end{eqnarray} 
   
   where,  $G_{IJKL}$ and $C_{I_{1}I_{2}I_{3}}$  are corresponded to the  gauge field $A$ and field strength $F$ \cite{b2}:
   
   \begin{eqnarray}
    && G_{IJKL}=-\frac{3}{\sqrt{2}}\frac{\kappa^{2}}{\lambda^{2}}\varepsilon(x^{11})F_{IJ}F_{KL}+... \nonumber\\&& \delta C_{ABC} =-\frac{\kappa^{2}}{6\sqrt{2}\lambda^{2}}\delta (x^{11}) tr \epsilon F_{AB}\nonumber\\&& G_{11ABC}=\partial_{11}C_{ABC}+....\nonumber\\&& F^{IJ}=\partial_{I}A^{J}-\partial_{J}A^{I}=\epsilon^{IJ}\partial_{I}A^{J} \label{s2}
    \end{eqnarray} 
    
   Here,  $\varepsilon(x^{11})$ is 1 for $x^{11}> 0$ and −1 for $x^{11}< 0$ and also $\delta(x^{11})=\frac{\partial \varepsilon}{\partial x^{11}}$. The gauge variation of the CGG-action, yields the following equation \cite{b2}:
    
    \begin{eqnarray}
      && \delta S_{CGG}|_{11}=-\frac{\sqrt{2}}{3456\bar{\kappa}^{2}}\int_{M^{11}}d^{11}x \varepsilon^{I_{1}I_{2}...I_{11}}\delta C_{I_{1}I_{2}I_{3}}G_{I_{4}...I_{7}}G_{I_{8}...I_{11}}= \nonumber\\&& - \frac{\bar{\kappa}^{4}}{128 \lambda^{6}}\int_{M^{10}}\epsilon^{I_{1}I_{2}..I_{10}} F_{I_{1}I_{2}}F_{I_{3}I_{4}}F_{I_{5}I_{6}}F_{I_{7}I_{8}}F_{I_{9}I_{10}}\label{s3}
      \end{eqnarray} 
     
    Above terms cancel the  anomaly of  ($S_{Bosonic-SUGRA}$) in eleven dimensional manifold \cite{b2}:
    
    \begin{eqnarray}
        && \delta S_{CGG}|_{11}=-\delta S_{Bosonic-SUGRA}=-S^{anomaly}_{Bosonic-SUGRA}\label{ss3}
        \end{eqnarray}

    Thus, to achieve to the  anomaly-free supergravity in eleven dimensions, we need to CGG terms. Now, we response to this question that what is the origin of these extra terms. In fact, we are investigating a theory that CGG terms are emerged in the action of supergravity without adding any by hand. To this end, we will use of higher dimensional supergravity theories. Also, we  select a unified shape for all fields by applying  Nambu-Poisson brackets and properties of string fields ($X$). We obtain \cite{h15,A1}:

    \begin{eqnarray}
     &&  X^{I_{i}}=y^{I_{i}}+A^{I_{i}}\nonumber\\&& \{ X^{I_{i}},X^{I_{j}} \}=\Sigma_{I_{i},I_{j}}\epsilon^{I_{i}I_{j}}\frac{\partial X^{I_{i}}}{\partial y^{I_{j}}}\frac{\partial X^{I_{j}}}{\partial y^{I_{j}}}= \nonumber\\&&\Sigma_{I_{i},I_{j}}\epsilon^{I_{i}I_{j}}\partial_{I_{i}}A^{I_{j}}=F^{I_{i}I_{j}}   \label{s4}
     \end{eqnarray} 
   
    Extending two dimensional brackets to four, we can derive the  form of GG-terms in supergravity in terms of scalar strings ($X$):
            
        \begin{eqnarray}
        && G_{IJKL}= \{ X^{I},X^{J},X^{K},X^{L} \} =\Sigma_{IJKL}\epsilon^{IJKL}\frac{\partial X^{I}}{\partial y^{I}}\frac{\partial X^{J}}{\partial y^{J}}\frac{\partial X^{K}}{\partial y^{K}}\frac{\partial X^{L}}{\partial y^{L}}\nonumber\\&&\Rightarrow \int d^{11}x\sqrt{g}\Big(G_{IJKL}G^{IJKL}\Big)=\nonumber\\&& \int d^{11}x\sqrt{g}\Big(\Sigma_{IJKL}\epsilon^{IJKL}\frac{\partial X^{I}}{\partial y^{I}}\frac{\partial X^{J}}{\partial y^{J}}\frac{\partial X^{K}}{\partial y^{K}}\frac{\partial X^{L}}{\partial y^{L}}\Sigma_{IJKL}\epsilon^{IJKL}\frac{\partial X^{I}}{\partial y^{I}}\frac{\partial X^{J}}{\partial y^{J}}\frac{\partial X^{K}}{\partial y^{K}}\frac{\partial X^{L}}{\partial y^{L}}\Big)  \label{s13}
        \end{eqnarray} 
          
     Above equation provides us this opportunity to use of more degrees of freedom and extract the CGG terms from GG-terms in supergravity. To this aime, we will add a three dimensional manifold to the eleven dimensional supergravity by applying the properties of scalar strings ($X$) in Nambu-Poisson brackets \cite{A1}:

         \begin{eqnarray}
         && X^{I}=y^{I}+A^{I}\Rightarrow\nonumber\\&& \frac{\partial X^{I_{5}}}{\partial y^{I_{5}}}\approx\delta ( y^{I_{5}})+.. \quad \frac{\partial X^{I_{6}}}{\partial y^{I_{6}}}\approx\delta ( y^{I_{6}})+.. \quad \frac{\partial X^{I_{7}}}{\partial y^{I_{7}}}\approx\delta ( y^{I_{7}})+...\nonumber\\&& \nonumber\\&&\int_{M^{N=3}}\rightarrow\int_{y^{I_{5}}+y^{I_{6}}+y^{I_{7}}}\frac{\partial X^{I_{5}}}{\partial y^{I_{5}}}\frac{\partial X^{I_{6}}}{\partial y^{I_{6}}}\frac{\partial X^{I_{7}}}{\partial y^{I_{7}}}=1+.. \label{s14}
         \end{eqnarray} 
         
      By adding three dimensional manifold of equation (\ref{s14}) to the eleven dimensional manifold of equation (\ref{s13}), we obtain:

          \begin{eqnarray}
          && \int_{M^{N=3}} \times \int_{M^{11}}\sqrt{g}\Big(G_{I_{1}I_{2}I_{3}I_{4}}G^{I_{1}I_{2}I_{3}I_{4}}\Big) = \nonumber\\&& \int_{M^{11}+y^{I_{5}}+y^{I_{6}}+y^{I_{7}}}\sqrt{g}\epsilon^{I_{4}I_{5}}\epsilon^{I_{4}I_{6}}\epsilon^{I_{5}I_{7}}\epsilon^{I_{6}I_{7}} G_{I_{1}I_{2}I_{3}I_{4}}G_{I_{1}I_{2}I_{3}I_{4}}\frac{\partial X^{I_{5}}}{\partial y^{I_{5}}}\frac{\partial X^{I_{5}}}{\partial y^{I_{6}}}\frac{\partial X^{I_{7}}}{\partial y^{I_{7}}}= \nonumber\\&&  \int_{M^{11}+M^{N=3}}\sqrt{g}CGG \nonumber\\&&\nonumber\\&&\Rightarrow C_{I_{5}I_{6}I_{7}}= \Sigma_{I_{5}I_{6}I_{7}}\epsilon^{I_{5}I_{6}I_{7}}\frac{\partial X^{I_{5}}}{\partial y^{I_{5}}}\frac{\partial X^{I_{5}}}{\partial y^{I_{6}}}\frac{\partial X^{I_{7}}}{\partial y^{I_{7}}}\label{s15}
          \end{eqnarray} 
          
      This equation provides three main results : 1. CGG terms can be emerged in the action of supergravity by adding a three dimensional manifold to eleven dimensinal supergravity. 2. 11-dimensional manifol + three-Lie-algebra=14-dimensional supergravity. 3. The form of C-terms is now clear in terms of string fields ($X^{i}$).
      
     To test the correctness of model we should re-derive the gauge variation of the CGG-action in equation (\ref{s3}) in terms of fields strenths . For this reason, using equation (\ref{s14} and \ref{s15}), we can calculate the gauge variation of C \cite{A1}:

           \begin{eqnarray}
           && X^{I}=y^{I}+A^{I}\Rightarrow \frac{\partial \delta_{A} X^{I}}{\partial y^{I}}=\delta ( y^{I}) \nonumber\\&&\nonumber\\&&\Rightarrow \int_{M^{N=3}+M^{11}}\delta_{A} C_{I_{5}I_{6}I_{7}}=\int_{M^{N=3}+M^{11}} \Sigma_{I_{5}I_{6}I_{7}}\epsilon^{I_{5}I_{6}I_{7}}\delta_{A}(\frac{\partial X^{I_{5}}}{\partial y^{I_{5}}}\frac{\partial X^{I_{5}}}{\partial y^{I_{6}}}\frac{\partial X^{I_{7}}}{\partial y^{I_{7}}})=\nonumber\\&& \int_{M^{N=3}+M^{10}}\Sigma_{I_{5}I_{6}}\epsilon^{I_{5}I_{6}}(\frac{\partial X^{I_{5}}}{\partial y^{I_{5}}}\frac{\partial X^{I_{6}}}{\partial y^{I_{6}}})=\nonumber\\&&\int_{M^{N=3}+M^{10}}F^{I_{5}I_{6}}\label{s16}
           \end{eqnarray}

                Using above result and   equation (\ref{s13})  we can obtain the gauge variation of the CGG action in equation of (\ref{s15}) \cite{A1}:
               
               \begin{eqnarray}
                    &&  \delta\int_{M^{11}+M^{N=3}}\sqrt{g}CGG = \nonumber\\&& \delta\int_{M^{11}+M^{N=3}}\sqrt{g}\epsilon^{I_{1}I_{2}I_{3}I_{4}I'_{1}I'_{2}I'_{3}I'_{4}I_{5}I_{6}I_{7}}\epsilon^{I_{5}I_{6}I_{7}}(\frac{\partial X^{I_{5}}}{\partial y^{I_{5}}}\frac{\partial X^{I_{6}}}{\partial y^{I_{6}}}\frac{\partial X^{I_{7}}}{\partial y^{I_{7}}}) G_{I_{1}I_{2}I_{3}I_{4}}G_{I'_{1}I'_{2}I'_{3}I'_{4}}= \nonumber\\&&\delta\int_{M^{11}+M^{N=3}}\sqrt{g}\epsilon^{I_{1}I_{2}I_{3}I_{4}I'_{1}I'_{2}I'_{3}I'_{4}I_{5}I_{6}I_{7}}\epsilon^{I_{5}I_{6}I_{7}}(\frac{\partial X^{I_{5}}}{\partial y^{I_{5}}}\frac{\partial X^{I_{6}}}{\partial y^{I_{6}}}\frac{\partial X^{I_{7}}}{\partial y^{I_{7}}}) \times \nonumber\\&&(\epsilon^{I_{1}I_{2}I_{3}I_{4}}\frac{\partial X^{I_{1}}}{\partial y^{I_{1}}}\frac{\partial X^{I_{2}}}{\partial y^{I_{2}}}\frac{\partial X^{I_{3}}}{\partial y^{I_{3}}}\frac{\partial X^{I_{4}}}{\partial y^{I_{4}}})(\epsilon^{I'_{1}I'_{2}I'_{3}I'_{4}}\frac{\partial X^{I'_{1}}}{\partial y^{I'_{1}}}\frac{\partial X^{I'_{2}}}{\partial y^{I'_{2}}}\frac{\partial X^{I'_{3}}}{\partial y^{I'_{3}}}\frac{\partial X^{I'_{4}}}{\partial y^{I'_{4}}})=\nonumber\\&& \int_{M^{10}+M^{N=3}}\sqrt{g}\epsilon^{I_{1}I_{2}I_{3}I_{4}I'_{1}I'_{2}I'_{3}I'_{4}I_{5}I_{6}I_{7}}\epsilon^{I_{5}I_{6}}(\frac{\partial X^{I_{5}}}{\partial y^{I_{5}}}\frac{\partial X^{I_{6}}}{\partial y^{I_{6}}}) \times \nonumber\\&&(\epsilon^{I_{1}I_{2}I_{3}I_{4}}\frac{\partial X^{I_{1}}}{\partial y^{I_{1}}}\frac{\partial X^{I_{2}}}{\partial y^{I_{2}}}\frac{\partial X^{I_{3}}}{\partial y^{I_{3}}}\frac{\partial X^{I_{4}}}{\partial y^{I_{4}}})(\epsilon^{I'_{1}I'_{2}I'_{3}I'_{4}}\frac{\partial X^{I'_{1}}}{\partial y^{I'_{1}}}\frac{\partial X^{I'_{2}}}{\partial y^{I'_{2}}}\frac{\partial X^{I'_{3}}}{\partial y^{I'_{3}}}\frac{\partial X^{I'_{4}}}{\partial y^{I'_{4}}})=\nonumber\\&& \int_{M^{10}+M^{N=3}}\sqrt{g}\epsilon^{I_{1}I_{2}I_{3}I_{4}I'_{1}I'_{2}I'_{3}I'_{4}I_{5}I_{6}I_{7}}(\epsilon^{I_{4}I_{5}}\frac{\partial X^{I_{4}}}{\partial y^{I_{4}}}\frac{\partial X^{I_{5}}}{\partial y^{I_{5}}})(\epsilon^{I'_{4}I_{6}}\frac{\partial X^{I'_{4}}}{\partial y^{I_{4}}}\frac{\partial X^{I_{6}}}{\partial y^{I_{6}}}) \times \nonumber\\&&(\epsilon^{I_{1}I_{2}}\frac{\partial X^{I_{1}}}{\partial y^{I_{1}}}\frac{\partial X^{I_{2}}}{\partial y^{I_{2}}})(\epsilon^{I'_{1}I'_{2}}\frac{\partial X^{I'_{1}}}{\partial y^{I'_{1}}}\frac{\partial X^{I'_{2}}}{\partial y^{I'_{2}}})(\epsilon^{I_{3}I'_{3}}\frac{\partial X^{I_{3}}}{\partial y^{I_{3}}}\frac{\partial X^{I'_{3}}}{\partial y^{I'_{3}}})= \nonumber\\&& \int_{M^{10}+M^{N=3}}\sqrt{g} F^{I_{1}I_{2}}F^{I'_{1}I'_{2}}F^{I'_{4}I_{6}}F^{I_{4}I_{5}}F^{I'_{3}I_{3}}\label{s17}
                    \end{eqnarray} 
                     
             The sign of these CGG -terms are reverse to the sign of  usual CGG-terms in equation (\ref{s3}) and thus, remove their effects and anomaly be removed completely.  On the other hand, applying  equations (\ref{s13} and \ref{s16})  we can rewrite the gauge variation of the CGG action in equation of (\ref{s15})\cite{A1}:
            
            \begin{eqnarray}
                 &&  \delta\int_{M^{11}+M^{N=3}}\sqrt{g}CGG = \nonumber\\&& \delta\int_{M^{11}+M^{N=3}}\sqrt{g}\epsilon^{I_{1}I_{2}I_{3}I_{4}I'_{1}I'_{2}I'_{3}I'_{4}I_{5}I_{6}I_{7}}\epsilon^{I_{5}I_{6}I_{7}}(\frac{\partial X^{I_{5}}}{\partial y^{I_{5}}}\frac{\partial X^{I_{5}}}{\partial y^{I_{6}}}\frac{\partial X^{I_{7}}}{\partial y^{I_{7}}}) G_{I_{1}I_{2}I_{3}I_{4}}G_{I'_{1}I'_{2}I'_{3}I'_{4}}= \nonumber\\&&\delta\int_{M^{11}+M^{N=3}}\sqrt{g}\epsilon^{I_{1}I_{2}I_{3}I_{4}I'_{1}I'_{2}I'_{3}I'_{4}I_{5}I_{6}I_{7}}\epsilon^{I_{5}I_{6}I_{7}}(\frac{\partial X^{I_{5}}}{\partial y^{I_{5}}}\frac{\partial X^{I_{5}}}{\partial y^{I_{6}}}\frac{\partial X^{I_{7}}}{\partial y^{I_{7}}}) \times \nonumber\\&&(\epsilon^{I_{1}I_{2}I_{3}I_{4}}\frac{\partial X^{I_{1}}}{\partial y^{I_{1}}}\frac{\partial X^{I_{2}}}{\partial y^{I_{2}}}\frac{\partial X^{I_{3}}}{\partial y^{I_{3}}}\frac{\partial X^{I_{4}}}{\partial y^{I_{4}}})(\epsilon^{I'_{1}I'_{2}I'_{3}I'_{4}}\frac{\partial X^{I'_{1}}}{\partial y^{I'_{1}}}\frac{\partial X^{I'_{2}}}{\partial y^{I'_{2}}}\frac{\partial X^{I'_{3}}}{\partial y^{I'_{3}}}\frac{\partial X^{I'_{4}}}{\partial y^{I'_{4}}})=\nonumber\\&& \int_{M^{10}+M^{N=3}}\sqrt{g}\epsilon^{I_{1}I_{2}I_{3}I_{4}I'_{1}I'_{2}I'_{3}I'_{4}I_{5}I_{6}I_{7}}\epsilon^{I_{5}I_{6}}(\frac{\partial X^{I_{5}}}{\partial y^{I_{5}}}\frac{\partial X^{I_{6}}}{\partial y^{I_{6}}}) \times \nonumber\\&&(\epsilon^{I_{1}I_{2}I_{3}I_{4}}\frac{\partial X^{I_{1}}}{\partial y^{I_{1}}}\frac{\partial X^{I_{2}}}{\partial y^{I_{2}}}\frac{\partial X^{I_{3}}}{\partial y^{I_{3}}}\frac{\partial X^{I_{4}}}{\partial y^{I_{4}}})(\epsilon^{I'_{1}I'_{2}I'_{3}I'_{4}}\frac{\partial X^{I'_{1}}}{\partial y^{I'_{1}}}\frac{\partial X^{I'_{2}}}{\partial y^{I'_{2}}}\frac{\partial X^{I'_{3}}}{\partial y^{I'_{3}}}\frac{\partial X^{I'_{4}}}{\partial y^{I'_{4}}})=\nonumber\\&& \int_{M^{10}+M^{N=3}}\sqrt{g}\epsilon^{I_{1}I_{2}I_{3}I_{4}I'_{1}I'_{2}I'_{3}I'_{4}I_{5}I_{6}I_{7}}(\epsilon^{I_{4}I_{5}}\frac{\partial X^{I_{4}}}{\partial y^{I_{4}}}\frac{\partial X^{I_{5}}}{\partial y^{I_{5}}})(\epsilon^{I'_{4}I_{6}}\frac{\partial X^{I'_{4}}}{\partial y^{I_{4}}}\frac{\partial X^{I_{6}}}{\partial y^{I_{6}}}) \times \nonumber\\&&(\epsilon^{I_{1}I_{2}I_{3}}\frac{\partial X^{I_{1}}}{\partial y^{I_{1}}}\frac{\partial X^{I_{2}}}{\partial y^{I_{2}}}\frac{\partial X^{I_{3}}}{\partial y^{I_{3}}})(\epsilon^{I'_{1}I'_{2}I'_{3}}\frac{\partial X^{I'_{1}}}{\partial y^{I'_{1}}}\frac{\partial X^{I'_{2}}}{\partial y^{I'_{2}}}\frac{\partial X^{I'_{3}}}{\partial y^{I'_{3}}})= \nonumber\\&& \int_{M^{10}+M^{N=3}}\sqrt{g} F^{2}F^{2}G^{3}G^{3}\label{s17}
                 \end{eqnarray} 
                  
            This result explains that the main reason of anomaly is joining fields in the bulk ($X$) to fields on the manifold and leading that their rank   grows  from 3 in Lie-3-algebra to $3+1$ ($G^{3} + X\rightarrow G^{3+1}$). For example, in 11-dimensional space-time with three algebra, extra fields in the bulk join to the fields on the manifold and leads that their rank grows  from three to four ($G^{3} + X\rightarrow G^{4}$).  In fact, if we work on the relative actions and add manifold of algebra to manifold of space-time, we can derive the real shape of fields and study their interactions. On the hand, an ($11+3$)dimensional manifold has the same physics of an ($11$)-dimensional space-time manifold inadditional to ($3$)-dimensional algebric manifold \cite{A1}. 
                                                                                                                                                                      
\section{Summary and conclusion }\label{o4}
 In this research, by extending  algebra from three to N and number of time-spacial  dimensions from 11 to M, we propose a new theory which in it, all dimensions and branes are created from nothing. We name this theory as G-theory.  In this theory, first, there is nothing. Then,  two types of degrees of freedoms with opposite signs are born which create two types of energies with opposite sign such as the sum over them be zero. These energies are excited, produce different dimensions and  lead to the emergence of two types of branes with opposite quantum numbers. On these branes, bosonic tensor fields live which their rank changes from zero to dimensions of brane and some of them play the role of graviton.  By compacting branes, fermionic superpartners of bosonic fields like the gravitino are created  and some dimensions take extra "i" factor and play the role of  time.  Gravitons produce bosonic wormholes which lead to the emergence of  one attractive potential between particles which grows and tends to infinity during contraction branch of branes. Thus,  attractive force between particles  leads to their closing, the end of expansion epoch and starting contraction branch.  Gravitinoes create fermionic wormholes which cause to the appearance of one repulisive potential between particles during expansion branch. Consequently,  the repulsive force between particles  leads to their getting away, the end of contraction epoch and begining expansion era.

\section*{Acknowledgments}
\noindent The work of Alireza Sepehri has been supported
financially by Research Institute for Astronomy and Astrophysics
of Maragha (RIAAM),Iran under research project No.1/4165-14.
The work was partly supported by VEGA Grant No. 2/0009/16.
R. Pincak would like to thank the
TH division in CERN for hospitality.



\begin{thebibliography}{99}
\bibitem{m1}M. J. Duff and K. S. Stelle, “Multimembrane solutions of D = 11 supergravity,”
Phys. Lett. B 253, 113 (1991).\\M. J. Duff  "M-history without the M", arXiv:1501.04098 [physics.hist-ph].\\
M.J. Duff (ed.)"The world in eleven-dimensions: Supergravity, supermembranes and M theory" (Michigan U.). 1999. 513 pp. Published in Bristol, UK:
IOP (1999) 513 p.
\bibitem{m2}M. J. Duff, P. S. Howe, T. Inami and K. S. Stelle, “Superstrings in D=10 from
Supermembranes in D=11,” Phys. Lett. B 191, 70 (1987).
\bibitem{m3}C. M. Hull and P. K. Townsend, “Unity of superstring dualities,” Nucl. Phys.
B 438, 109 (1995) [hep-th/9410167].
\bibitem{m4}P. K. Townsend, “The eleven-dimensional supermembrane revisited,” Phys.
Lett. B 350, 184 (1995) [hep-th/9501068].
\bibitem{t1} Edward Witten, “String theory dynamics in various dimensions,” Nucl. Phys. B
443, 85 (1995) [hep-th/9503124].
\bibitem{h13} J. Bagger and N. Lambert, Gauge Symmetry and Supersymmetry of Multiple M2-Branes,
Phys. Rev. D 77, 065008 (2008) [arXiv:0711.0955 [hep-th]].

\bibitem{h14} A. Gustavsson, Algebraic structures on parallel M2-branes, arXiv:0709.1260 [hep-th]; A.Sepehri, R.Pincak, in proceeding.

\bibitem{h15} Pei-Ming Ho, Yutaka Matsuo, JHEP 0806:105,2008.

\bibitem{h16}Sunil Mukhi, Constantinos Papageorgakis,
JHEP 0805:085,2008.

\bibitem{h1}Ahmed Farag Ali and Saurya Das, Phys.Lett. B741 (2015) 276.

\bibitem{h2}Alireza Sepehri, Physics Letters B 748 (2015) 328335,
arXiv:1508.01407 [gr-qc].

\bibitem{h3}Alireza Sepehri, Phys. Lett. A, 380 (2016) 2247.

\bibitem{h4}Alireza Sepehri, Farook Rahaman, Salvatore Capozziello, Ahmed Farag Ali, Anirudh Pradhan, 
Eur.Phys.
J. C76 (2016) no.5, 231.

\bibitem{h5}Alireza Sepehri, Mohammad Reza Setare, Salvatore Capozziello, Eur.Phys.J. C75 (2015) no.12, 618.
\bibitem{h6}Alireza Sepehri, Anirudh Pradhan, A. Beesham, Jaume de Haro, Phys.Lett. B760 (2016) 94-100.
\bibitem{h7}Alireza Sepehri etal., PhysicsLetters B747(2015)1-8
 \bibitem{h8} A. Sepehri, F. Rahaman, A. Pradhan and I. H. Sardar, Phys. Lett. B 741, 92 (2014).

\bibitem{h9}Claudia de Rham, Andrew J. Tolley, Shuang-Yong Zhou, JHEP 1604 (2016) 188.\\ Miguel Cruz, Efraín Rojas,  Class. Quantum Grav. 30 (2013) 115012. \\Alireza Sepehri, Richard Pincak,The birth of the universe in a new G-Theory approach, in proceeding.

\bibitem{h10} R.C. Myers, JHEP 9912, 022 (1999), hep-th/9910053.
\bibitem{h11} Neil R. Constable, Robert C. Myers, Oyvind Tafjord
, JHEP 0106:023,2001; A.A. Tseytlin, hep-th/9908105.
\bibitem{h12}Chong-Sun Chu, Douglas J Smith, JHEP 0904:097,2009.

\bibitem{r12}J.Kluson, JHEP 0011 (2000) 016.

\bibitem{rr12}J. Kluson, Phys.Rev. D64 (2001) 126006  

\bibitem{h17}S. Capozziello, V.F. Cardone, A. Troisi, Phys.Rev. D71 (2005) 043503.
\bibitem{h18}S. Capozziello, V.F. Cardone, V. Salzano, Phys.Rev.D78:063504,2008.
\bibitem{h19}Gianluca Grignani, Troels Harmark, Andrea Marini, Niels A. Obers, Marta Orselli, JHEP 1106:058,2011.
 
\bibitem{h20}Mahdi Kord Zangeneh, Francisco S. N. Lobo, Nematollah Riazi,  Phys.Rev.D90:024072,2014. 

\bibitem{b1}Petr Horava, Edward Witten, Nucl. Phys. B460 (1996) 506, hep-th/9510209.
\bibitem{b2}Petr Horava, Edward Witten, Nucl.Phys.B475:94-114,1996.
\bibitem{A1}Alireza Sepehri, Anomaly cancellation in D+N dimensional supergravity, submitted to journal, in proceeding.
\end{thebibliography}
\end{document}